\documentclass[preprint2,times]{aastex61} 
\usepackage[utf8]{inputenc}

\newcommand{\fblue}[1]{F$_{70}$\,}
\newcommand{\fgreen}[1]{F$_{100}$\,}
\newcommand{\fred}[1]{F$_{160}$\,}
\newcommand{\tiunit}{$\mathrm{J\,m^{-2}\,s^{-1/2}\,K^{-1}}$}
\newcommand{\chired}{$\chi^2_{\rm r}$}
\def\absmag{H$_\mathrm{V}$}

\def\geomalb{$\mathrm p_\mathrm{V}$}

\shortauthors{A. Farkas-Tak\'acs et al.}
\shorttitle{Properties of the irregular satellite system around Uranus}
\begin{document}
\sloppy

%\title{Characteristics of Uranian irregular satellites from K2, Herschel and Spitzer observations}
\title{Properties of the irregular satellite system around Uranus inferred from K2, Herschel\footnote{Herschel is an ESA space observatory with science instruments provided by European-led Principal Investigator consortia and with important participation from NASA.} and Spitzer observations}
\correspondingauthor{A. Farkas-Tak\'acs}
\email{farkas.aniko@csfk.mta.hu}

\author{A. Farkas-Tak\'acs}
\affiliation{Konkoly Observatory, Research Centre for Astronomy and Earth Sciences, Hungarian Academy of Sciences, Konkoly Thege Mikl\'os \'ut 15-17, H-1121 Budapest, Hungary}

\author{Cs. Kiss}
\affiliation{Konkoly Observatory, Research Centre for Astronomy and Earth Sciences, Hungarian Academy of Sciences, Konkoly Thege Mikl\'os \'ut 15-17, H-1121 Budapest, Hungary}

\author{A. P\'al}
\affiliation{Konkoly Observatory, Research Centre for Astronomy and Earth Sciences, Hungarian Academy of Sciences, Konkoly Thege Mikl\'os \'ut 15-17, H-1121 Budapest, Hungary}
\affiliation{Department of Astronomy, Lor\'and E\"otv\"os University, 
	P\'azm\'any P\'eter s\'et\'any 1/A, 
	1117 Budapest, Hungary}

\author{L. Moln\'ar}
\affiliation{Konkoly Observatory, Research Centre for Astronomy and Earth Sciences, Hungarian Academy of Sciences, Konkoly Thege Mikl\'os \'ut 15-17, H-1121 Budapest, Hungary}

\author{Gy. M. Szab\'o}
\affiliation{Konkoly Observatory, Research Centre for Astronomy and Earth Sciences, Hungarian Academy of Sciences, Konkoly Thege Mikl\'os \'ut 15-17, H-1121 Budapest, Hungary}
\affiliation{ELTE E\"otv\"os Lor\'and University, Gothard Astrophysical Observatory, Szombathely, Hungary}	
\author{O. Hanyecz}
\affiliation{Konkoly Observatory, Research Centre for Astronomy and Earth Sciences, Hungarian Academy of Sciences, Konkoly Thege Mikl\'os \'ut 15-17, H-1121 Budapest, Hungary}
\affiliation{Department of Astronomy, Lor\'and E\"otv\"os University, 
	P\'azm\'any P\'eter s\'et\'any 1/A, 
	1117 Budapest, Hungary}

\author{K. S\'arneczky}
\affiliation{Konkoly Observatory, Research Centre for Astronomy and Earth Sciences, Hungarian Academy of Sciences, Konkoly Thege Mikl\'os \'ut 15-17, H-1121 Budapest, Hungary}

\author{R. Szab\'o}
\affiliation{Konkoly Observatory, Research Centre for Astronomy and Earth Sciences, Hungarian Academy of Sciences, Konkoly Thege Mikl\'os \'ut 15-17, H-1121 Budapest, Hungary}

\author{G. Marton}
\affiliation{Konkoly Observatory, Research Centre for Astronomy and Earth Sciences, Hungarian Academy of Sciences, Konkoly Thege Mikl\'os \'ut 15-17, H-1121 Budapest, Hungary}

\author{M. Mommert}
\affiliation{Department of Physics and Astronomy, Northern Arizona  
    University, PO Box 6010, Flagstaff, AZ 86011, USA}

\author{R. Szak\'ats}
\affiliation{Konkoly Observatory, Research Centre for Astronomy and Earth Sciences, Hungarian Academy of Sciences, Konkoly Thege Mikl\'os \'ut 15-17, H-1121 Budapest, Hungary}

\author{T. M\"uller}
\affiliation{Max-Plank-Institut f\"ur extraterrestrsiche Pyhsik, Garching, Germany}

\author{L.L. Kiss}
\affiliation{Konkoly Observatory, Research Centre for Astronomy and Earth Sciences, Hungarian Academy of Sciences, Konkoly Thege Mikl\'os \'ut 15-17, H-1121 Budapest, Hungary}
\affiliation{Sydney Institute for Astronomy, School of Physics A28, 
	University of Sydney, 
	NSW 2006, Australia}

\date{\today}

%\author{%
%A. Farkas-Tak\'acs\inst{1} \and 
%Cs. Kiss\inst{1} \and
%A. P\'al\inst{1,2} \and
%L. Moln\'ar\inst{1} \and
%Gy. M. Szab\'o\inst{1,2} \and
%L. L. Kiss\inst{2,4} \and
%O. Hanyecz\inst{1,2} \and
%K. S\'arneczky\inst{1} \and
%R. Szab\'o\inst{1} \and
%G. Marton\inst{1} \and
%M. Mommert\inst{5} \and 
%R. Szak\'ats\inst{1} \and 
%T. M\"uller\inst{6}
%}

%\institute{%
%Konkoly Observatory, Research Centre for Astronomy and Earth Sciences, Hungarian Academy of Sciences, Konkoly-Thege Mikl\'os \'ut 15-17, H-1121 Budapest, Hungary; e-mail: \texttt{farkas.aniko@csfk.mta.hu} \and
%Department of Astronomy, Lor\'and E\"otv\"os University, 
%	P\'azm\'any P\'eter s\'et\'any 1/A, 
%	1117 Budapest, Hungary \and
%ELTE Gothard Astrophysical Observatory, 
%	9704 Szombathely, Szent Imre herceg \'utja 112, Hungary \and	
%Sydney Institute for Astronomy, School of Physics A28, 
%	University of Sydney, 
%	NSW 2006, Australia \and
%Department of Physics and Astronomy, Northern Arizona  
%   University, PO Box 6010, Flagstaff, AZ 86011, USA \and
%    Max-Plank-Institut f\"ur extraterrestrsiche Pyhsik, Garching, Germany
%}

\begin{abstract}
In this paper we present visible range light curves of the irregular Uranian satellites Sycorax, Caliban, Prospero, Ferdinand and Setebos taken with Kepler Space Telescope in the course of the K2 mission. Thermal emission measurements obtained with the Herschel/PACS and Spitzer/MIPS instruments of Sycorax and Caliban were also analysed and used to determine size, albedo and surface characteristics of these bodies. We compare these properties with the rotational and surface characteristics of irregular satellites in other giant planet systems and also with those of main belt and Trojan asteroids and trans-Neptunian objects. Our results indicate that the Uranian irregular satellite system likely went through a more intense collisional evolution than the irregular satellites of Jupiter and Saturn. Surface characteristics of Uranian irregular satellites seems to resemble the Centaurs and trans-Neptunian objects more than irregular satellites around other giant planets, suggesting the existence of a compositional discontinuity in the young Solar system inside the orbit of Uranus. 
\end{abstract}

%\maketitle

\section{Introduction}

Giant planets possess basically two distinct types of satellites concerning orbital dynamics. \emph{Regular} satellites are characterized by orbits with small eccentricity that are very close to the planets' equatorial plane with always prograde orbit within $\sim0.05\,r_H$, where $r_H$ is the radius of Hill sphere of host planet. In contrast, \emph{irregular} satellites have moderate-to-high eccentricities and inclinations with prograde or retrograde orbits up to $0.65\,r_H$ from their host planets.
The existence of two classes of satellites reflects two different ways of evolution: regular satellites likely formed in the same subnebula as the host planet while irregular satellites could not have formed at their present orbits. The most accepted scenario currently is that they have been captured from the inside of the planet's subnebula in the last phase of planet formation on temporarily orbits, then settled through some kind of loss of angular momentum \citep[see][for a review]{Nicholson}. 
%some loss of energy is needed to make the capture permanent. There are several possible ways: (1) The mass of the planets increases through the accretion causing change in the radius of the Hill sphere \citep{Heppe}. (2) The gas drag through the evolving disk around the planet  \citep{Pollack}. (3) Collisions or close encounters to a preexisting satellite or another temporarily captured planetesimal \citep{Colombo}.

There were several deep surveys of irregular satellites in the 2000s that established the basis of the currently known set of irregular satellites \citep{Gladman02,Gladman01,Holman03,Sheppard03,SheppardIAUC,Sheppard05, Sheppard06}. These surveys provided the main orbital characteristics of the satellites found and allowed the identification of orbital grouping/families around the specific planets \citep{Nicholson}.

Unlike the members of other small body populations in the Solar System, irregular satellites may have remained close to their formation locations and their compositions may be an intermediate one between that of the main belt asteroids and the icy trans-Neptunian objects. 
The physical characterization of these satellites is, however, still a challenging task due to the large distance and the typical size below 100\,km, even in the case of the closest Jovian system. Among these characteristics, light curves provide information on the shape and/or surface albedo variegations and may give hints of the internal structure and strength in the case of fast rotators; the distribution of rotational frequencies and amplitudes are important properties of a small body population \citep{Pravec2002}. High-quality light curve is available for the largest Jovian irregular, Himalia \citep{Pilcher}, rotational properties are known for the Jovian satellites Lysithea, Ananke, Carme and Sinope \citep{Luu} and the Cassini spacecraft provided rotation periods for many irregular satellites in the Saturnian system \citep{Denk2013,Denk2014,Denk2015}. 
In the Uranian system, \citet{Maris01,Maris07} performed the investigation of the light curves of irregular satellites and obtained rotational characteristics for Sycorax, Caliban, Prospero and Setebos. However, in these latter cases the results are based on sparsely sampled data due to the observing capabilities of the telescopes and the large distance (hence faintness) of these satellites.  

Broad-band colors are the most readily available tools that can be used to characterize the surface of the irregular satellites \citep[see][and references therein]{Nicholson}. In the Jovian system the colors are similar to those of carbonaceous asteroids and Jovian Trojans, while in the Saturnian system they show somewhat redder surfaces. However, in both systems the colors are still far from that of the red material typically found in the Kuiper belt. In the Uranian system the colors show a wide variety, but there are certainly satellites that show typical ``Kuiper belt" colors \citep{Maris01,Maris07,Romon01,Grav04}. 

Well-established size and albedo values are available for a limited sample only, mainly from data of space probes -- e.g. Himalia and Phoebe by Cassini \citep{Porco03,Porco05}, or Nereid by Voyager-2 \citep{Thomas1991}. Recently, thermal emission data obtained with the MIPS camera of the Spitzer Space Telescope \citep{MIPS} and the PACS camera of the Herschel Space Observatory \citep{PACS} also provided independent size and albedo estimates for Sycorax \citep{Lellouch13} and Nereid \citep{Kiss2016}. 

As we noted above, ground-based observations could not place strong constraints on the rotation of most of the irregular satellites observed earlier. However, it is demonstrated in recent works \citep{Pal15,Pal16,Szabo2016,Szabo2017} that data from the extended \textit{Kepler} mission \citep[K2]{Howell14} can be very effectively used to obtain rotational light curves of Solar system bodies due to uninterrupted photometric time series of several tens of days in length, including main belt and Jovian Trojan asteroids, and trans-Neptunian objects, even at the brightness level of typical irregular satellites. Lately, a thorough light curve analysis of the Neptunian irregular satellite Nereid was also performed \citep{Kiss2016}, showing the great capabilities of K2 measurements for this kind of applications. 

%In the color-color (B--V, V--R) diagram of irregular satellites of Uranus and Neptune two groups separated: some of neutral color and the slightly red. Furthermore, the irregular satellites of Uranus were classified into two possible dynamic family: one of them the Caliban family and the other one is the Sycorax family \citep{Grav04}.

%By these, the  dynamic study of irregular satellites and knowledge of the physical and chemical properties of them can provide information on the origins and developments in them which is associated with the formation of the solar system.

%In the case of Uranus, \citet{Sheppard05} performed an ultra-deep survey looking for irregular satellites and detected moons down to the limiting magnitude of m$_{\mathrm R}$\,=\,26.1\,mag that corresponds to a $\sim$7\,km diameter assuming a very dark surface. 

In this paper we present the results of Uranian irregular satellite observations performed with the \textit{Kepler} space telescope in Campaign 8 of the K2 mission. We provide light curves and derive rotational characteristics for Sycorax, Caliban, Prospero, Setebos and Ferdinand (Sect.~\ref{sect:rotation}). In addition, we use thermal emission measurements of the Spitzer Space Telescope and the Herschel Space Observatory to derive more accurate size and albedo for Sycorax; we also give constraints on the size and albedo of Caliban, Trinculo and Ferdinand (Sect.~\ref{sect:thermal}) based on Herschel/PACS observations. Our results are compared with the properties of other irregular satellites and other small body populations of our Solar system (Sects.~\ref{sect:rotcomp} and \ref{sect:albedocolor}). 

\section{Observations}

\subsection{Kepler/K2 measurements}
Continuous photometry from the \textit{Kepler} space telescope may provide accurate and unbiased rotation rates and amplitudes for solar system targets. The telescope gathers light in a wide visual band spanning from 420 to 900 nm, and follows a step--and--stare scheme in the K2 mission, observing different fields for up to 80 days along the Ecliptic plane \citep{Howell14}. \textit{Kepler} observed Uranus and its vicinity during Campaign 8 of the mission for 78.73 days, between 2016 January 4.55 and March 23.28. Apart from the planet, four irregular moons, Caliban, Setebos, Sycorax, and Prospero, were also proposed and selected for observations (GO8039, PI: A. P\'al)\footnote{https://keplerscience.arc.nasa.gov/k2-approved-programs.html}. 

We applied the same pipeline for the reduction of the \textit{Kepler} observations that we used in previous works, e.g., to determine the light variations and rotation rates of various Trans-Neptunian Objects, main-belt and Trojan asteroids, or of the moon Nereid \citep{Kiss2016,Pal15,Pal16,Szabo2016,Szabo2017}. The method is based on the FITSH\footnote{http://fitsh.net} software package \citep{Pal2012}: the processing steps are detailed in the previous papers and in the companion paper that describes the light curves of main-belt asteroids drifting though the same image mosaic \citep{Molnar2017}, also providing information on the limitations and capabilities. In short, we created mosaic images from the individual Target Pixel Files (TPFs) which contain the time-series photometric information (time, flux, flux error, background) for each downloaded pixel around a given target, as opposed to light curve files that contain a pipeline-extracted brightness summed in an aperture as a function of time. For more details the reader is directed to Kepler Archive Manual\footnote{https://archive.stsci.edu/kepler/manuals/archive\_manual.pdf}. We then derived the astrometric solutions for the mosaic images, using the USNO-B1.0 catalog \citep{USNO}, where the K2 full-frame images from the campaign were exploited as initial hints for the source cross matching. Then, we registered the images into the same reference system,  
and subtracted a median image from each image. 
This median image was created from a selection of individual images that did not include the obvious diffractions pattern contamination from Uranus. We then applied aperture photometry at the positions of the satellites. The sharp images of the stars that were shifted to compensate the attitude changes of the telescope create characteristic residuals in the differential
images that may contaminate our photometry. Therefore we filtered out the epochs when the scatter of the background pixels in the photometric annulus was high. The per-cadence photometric uncertainty values were derived from the shot noise of Kepler and from the estimated background
noise.

\begin{figure*}
\includegraphics[width=\textwidth]{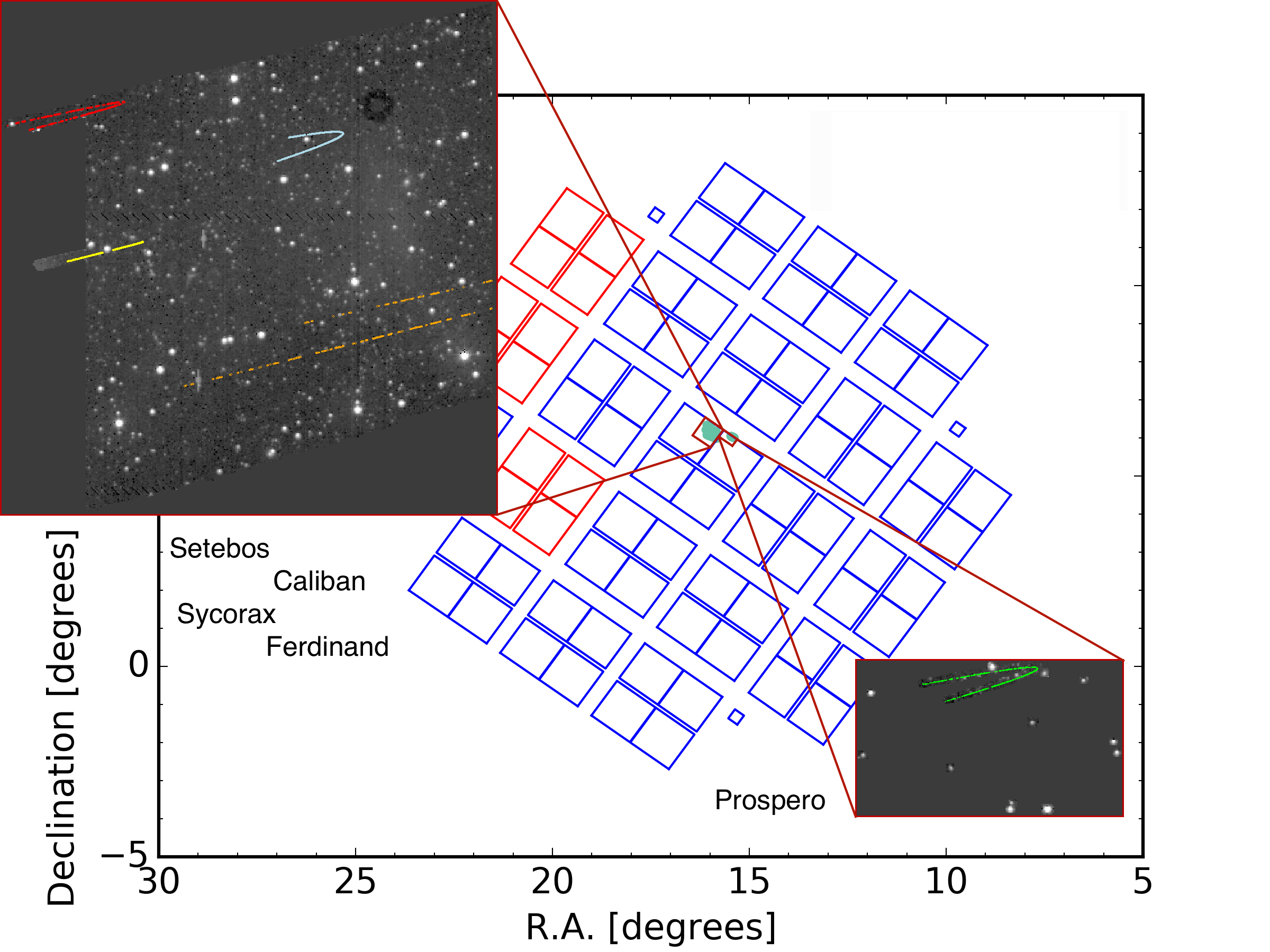}
\caption{The median image of the Uranus mosaic we used for differential image photometry, plus the vicinity of Prospero on the adjacent CCD module, as seen by \textit{Kepler}, overplotted on the K2 Campaign 8 field-of-view. Colored dots show the positions of the satellites at each successful photometric measurement. Red: Setebos; yellow: Sycorax; light blue: Caliban; orange: Ferdinand; green: Prospero. \label{fig:k2map}}
\end{figure*}

All observations were collected in long cadence mode, with a sampling of 29.4 min. Three of the proposed satellites fell into, or near the large mosaic that also covered the apparent motion of Uranus (Fig.~\ref{fig:k2map}). Other satellites were also present in the same mosaic, and we successfully detected the light variations of a fourth one, Ferdinand. 

\begin{figure}
\includegraphics[width=\columnwidth]{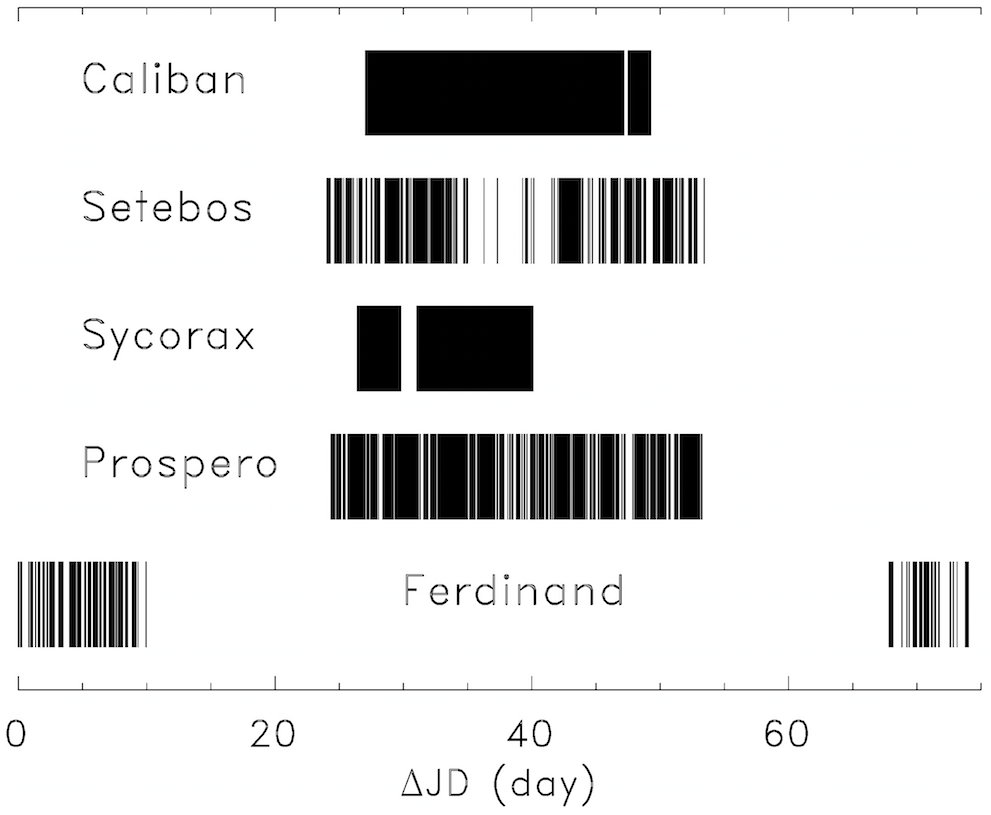}
\caption{Temporal distribution of data points used for photometry and light curve construction for Caliban, Setebos, Sycorax, Prospero and Ferdinand. $\Delta$JD is the date from the start of Campaign 8. \label{fig:sampling}}
\end{figure}

Prospero fell onto an adjacent CCD module, and its motion was covered with a narrow band of pixels. In that case we collected the Target Pixel Files of nearby, unrelated targets into a small mosaic around the track of Prospero in order to have a good astrometric solution. 
The log of observations is summarized in Table~\ref{tab:keplerlog}. The duty cycle there shows the ratio of accepted photometric data points to the number of cadences when the satellites were present on the images. We note that Ferdinand was observed twice, at the beginning and the end of the campaign: the two rows in the table refer to the entire length of the observation and the sections when Ferdinand was on the Kepler CCD array, respectively. The temporal distribution of data points we used for photometry is presented in Fig.~\ref{fig:sampling}. 
We also checked other satellites that fell on the CCD array, e.g., Stephano (24.5 mag) or Trinculo (25.5 mag), but found no meaningful signals in their photometry. 
%%%%
\begin{table}
\begin{tabular}{lccccc}
\hline
Name & Start & Length & Points & Duty & K$_p$ \\
~ & TBJD (d) & d & ~ & cycle & mag \\
\hline
Caliban & 7419.16 & 22.23 & 1019 & 0.93 & 21.99 \\
Setebos & 7416.15 & 28.85 & 495 & 0.35 & 22.87 \\
Sycorax & 7418.55 & 13.67 & 599 & 0.89 & 20.18 \\
Prospero & 7416.48 & 28.89 & 793 & 0.56 & 22.94 \\
Ferdinand & 7392.12 & 74.01 & 270 & 0.07 & 23.12 \\
\textit{on array} & ~ & 16.14 &  ~ & 0.32 &  ~ \\
\hline
\end{tabular}
\caption{Log of the \textit{Kepler} observations. TBJD means truncated BJD, BJD-2450000. The last line shows the values for Ferdinand if we ignore the 57.87 d long period it spent off the detector array during the campaign. Duty cycle  is 1.0 when all photometric points are retained and 0 when all had to be discarded. K$_p$ is the mean brightness of the target in the Kepler photometric system.}
\label{tab:keplerlog}
\end{table}
%%%%
%%%%
%%%%

\subsection{Infrared data}

\subsubsection{Sycorax}

Sycorax was observed with the MIPS camera of the Spitzer Space Telescope \citep{MIPS} at two epochs, on December 27 and 29, 2008, both at 24 and 70\,$\mu$m. The summary of these observations is given in Table~\ref{table:thermal} below. Sycorax was successfully detected at both epochs and at both wavelengths. We used the same data reduction and photometry pipeline as in \citet{stansberry2008,stansberry2012}. 
The MIPS instrument team data analysis tools \citep{Gordon2005} were used to produce flux-calibrated images for each band, and the contribution of background objects were subtracted \citep[see][]{stansberry2008}. Aperture photometry was performed both on the original and final images and the final
flux values were obtained using the aperture corrections by \citet{Gordon2007} and \citet{Engelbracht2007}. Color correction of the in-band fluxes were done following \citep{Stansberry2007}. 
The flux densities obtained are presented in Table~\ref{table:thermal2}. 

\begin{figure}[ht!]
\includegraphics[width=8.5cm]{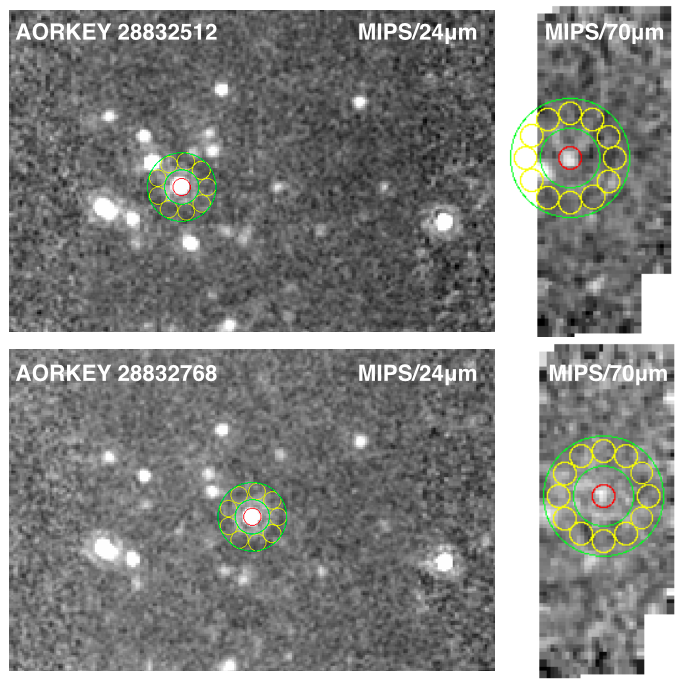}
\caption{Spitzer/MIPS 24 and 70\,$\mu$m images of Sycorax. The top and bottom rows present the images corresponding to the AORKEYs 28832512 and 28832768, respectively. On each image, red circles mark the target aperture and yellow apertures mark the positions used for background determinations \citep[see][for details]{stansberry2008,stansberry2012} \label{fig:mipsall}}
\end{figure}
\begin{table*}
\begin{tabular}{cccccccc}
\hline
Instrument & AORKEY/ & Target & Band & Date      & r    & $\Delta$ & $\alpha$ \\
           &    OBSID    &        & ($\mu$m) & (JD)  & (AU) & (AU) & (deg)     \\
\hline
Spitzer/MIPS & 28832512  & Sycorax & 24 & 2454827.702 & 20.081 & 19.672 & 2.68 \\
             & 28832512  &         & 70 & 2454827.718 & 20.081 & 19.672 & 2.68 \\
             & 28832768  &         & 24 & 2454829.672 & 20.081 & 19.704 & 2.71 \\
             & 28832768  &         & 70 & 2454829.688 & 20.081 & 19.704 & 2.71 \\
\hline             
Herschel/PACS & 1342221837-38 & Sycorax & 70 & 2455710.589 & 20.084 & 20.519 & 2.62 \\
             & 1342221875-76   &    &   70   & 2455710.939 & 20.084 & 20.514 & 2.62 \\
             &1342221839-40 &       &   100  & 2455710.617 & 20.084 & 20.519 & 2.62 \\
             & 1342221877-78 &      &   100  & 2455710.966 & 20.084 & 20.513 & 2.63 \\
             & 1342221837-40 &      &   160  & 2455710.589 & 20.084 & 20.519 & 2.62 \\
             & 1342221875-78 &      &   160  & 2455710.939 & 20.084 & 20.514 & 2.62 \\
\hline
Herschel/PACS & 1342236891-92 & Caliban & 70 & 2455933.979 & 20.119 & 20.359 & 2.73 \\
              & 1342237436-37 &         & 70 & 2455940.001 & 20.117 & 20.457 & 2.63 \\
			  & 1342236891-92 & 	    & 160 & 2455933.979 & 20.119 & 20.359 & 2.73 \\
              & 1342237436-37 &         & 160 & 2455940.001 & 20.117 & 20.457 & 2.63 \\
\hline
Herschel/PACS & 1342236891-92 & Trinculo & 70 & 2455933.979 & 20.051 & 20.293 & 2.73 \\
              & 1342237436-37 &          & 70 & 2455940.001 & 20.048 & 20.390 & 2.64 \\
		 	  & 1342236891-92 & 		 & 160 & 2455933.979 & 20.051 & 20.293 & 2.73 \\
              & 1342237436-37 &          & 160 & 2455940.001 & 20.048 & 20.390 & 2.64 \\
\hline
Herschel/PACS & 1342236891-92 & Ferdinand & 70 & 2455933.979 & 19.930 & 20.174 & 2.75 \\
              & 1342237436-37 &          & 70 & 2455940.001 & 19.929 & 20.173 & 2.65 \\
			  & 1342236891-92 & 	     & 160 & 2455933.979 & 19.930 & 20.174 & 2.75 \\
              & 1342237436-37 &          & 160 & 2455940.001 & 19.929 & 20.173 & 2.65 \\
\hline
\end{tabular}
\caption{Summary of thermal infrared observations. The columns are: (1) Telescope/instrument; (2) AORKEY (Spitzer/MIPS) or OBSID (Herschel/PACS); 
(3) target -- in the case of Caliban the target was caught in a larger, off-Uranus field;
(4) band (nominal wavelength in $\mu$m); 
(5) date of the observation (Julian date); 
(6) heliocentric distance; 
(7) distance from observer; 
(8) phase angle; 
\label{table:thermal}}
\end{table*}

%%%%%%%%%%%%%%%%%%%%%%%%%%%%%%%%%%%%%%%%%%%%%%%%%%%%%%%%%%%
\begin{table*}
\begin{tabular}{ccccccc}
\hline
Target & Detector/ & $\lambda_{\mathrm{eff}}$ & F$_i$ 	& $C_{\lambda}$	& $F_m$		& H$_\mathrm{V/R}$ \\
       & filter	   &    ($\mu$m)    	 & (mJy) 	    & 			  	& mJy		 &  (mag) \\
\hline
Sycorax & MIPS 24 &  23.68 & 3.017$\pm$0.045 & 0.96$\pm$0.01 & 3.14$\pm$0.16 & 7.50$\pm$0.04 \\
        & MIPS 70 &  71.42 & 14.68$\pm$2.78 & 0.92$\pm$0.01 & 16.07$\pm$2.89 &  (Grav et al., 2004)\\
        & MIPS 24 &  23.68 & 3.109$\pm$0.046 & 0.96$\pm$0.01 &  3.17$\pm$0.17 &  \\
        & MIPS 70 &  71.42 & 19.12$\pm$2.70 & 0.92$\pm$0.01 & 20.78$\pm$3.11 &  \\           
		& PACS 70 &  70.0  & 16.7$\pm$0.6 & 0.98$\pm$0.01 & 17.0$\pm$1.0 &  \\
        & PACS 100&  100.0 & 15.3$\pm$1.6 & 1.00$\pm$0.01 & 15.3$\pm$1.8 &  \\
        & PACS 160&  160.0 & 5.3$\pm$3.1 & 1.04$\pm$0.01 &  5.5$\pm$3.2 &  \\
\hline
Caliban & PACS 70 &  70.0  & 1.4$\pm$0.8 & 0.98$\pm$0.01 & 1.4$\pm$0.8 & 9.16$\pm$0.016 \\
		& PACS 160&  160.0 & $<$3\,mJy & 1.04$\pm$0.01 & $<$3\,mJy & (Grav et al., 2004) \\
\hline
Trinculo& PACS 70 &  70.0  & $<$0.8\,mJy & 0.98$\pm$0.01 & $<$0.8\,mJy & 11.92$\pm$0.18 \\
		& PACS 160&  160.0 & $<$3\,mJy & 1.04$\pm$0.01  & $<$3\,mJy & (Grav et al., 2004) \\
\hline
Ferdinand& PACS 70 & 70.0  & $<$0.8\,mJy & 0.98$\pm$0.01 & $<$0.8\,mJy & 12.5$\pm$0.1(R) \\
		 & PACS 160& 160.0 & $<$3\,mJy & 1.04$\pm$0.01  & $<$3\,mJy &  (this work)\\
\hline              
\end{tabular}
\caption{Summary of thermal infrared observations. The columns are: 
(1) target -- in the case of Caliban the target was caught in a larger, off-Uranus field;
(2) detector and filter combination; 
(3) effective wavelength of the band ($\lambda _{eff}$); 
(4) in-band flux density (F$_i$); 
(5) colour correction factor (C$_\lambda$); 
(6) monochromatic flux (F$_m$\,=\,F$_i$/C$_\lambda$); 
(7) absolute magnitude in V or R band; Herschel/PACS flux densities of Sycorax are taken from \citet{Lellouch13}
\label{table:thermal2}}
\end{table*}
%%%%%%%%%%%%%%%%%%%%%%%%%%%%%%%%%%%%

Sycorax was also observed in dedicated observations with the PACS camera of the Herschel Space Observatory, in the framework of the `TNOs are Cool!' Herschel Open Time Key Program \citep{Muller2009}. The flux densities derived from these observations have already been presented in \citet{Lellouch13}. 

We used both the Spitzer/MIPS and Herschel/PACS data for modeling of the thermal emission of the satellite (see Sect.~\ref{sycoraxthermal}). 

%%%%%%%%%%%%%%%%%%%%%%%%%%%%%%%%%%%%%%%%%%%%%%%%%%%%%%%%
\subsubsection{Serendipitous Herschel/PACS observations of irregular satellites}

%%%%%%%%%%%%%%
\begin{figure}[ht!]
\includegraphics[width=8.5cm]{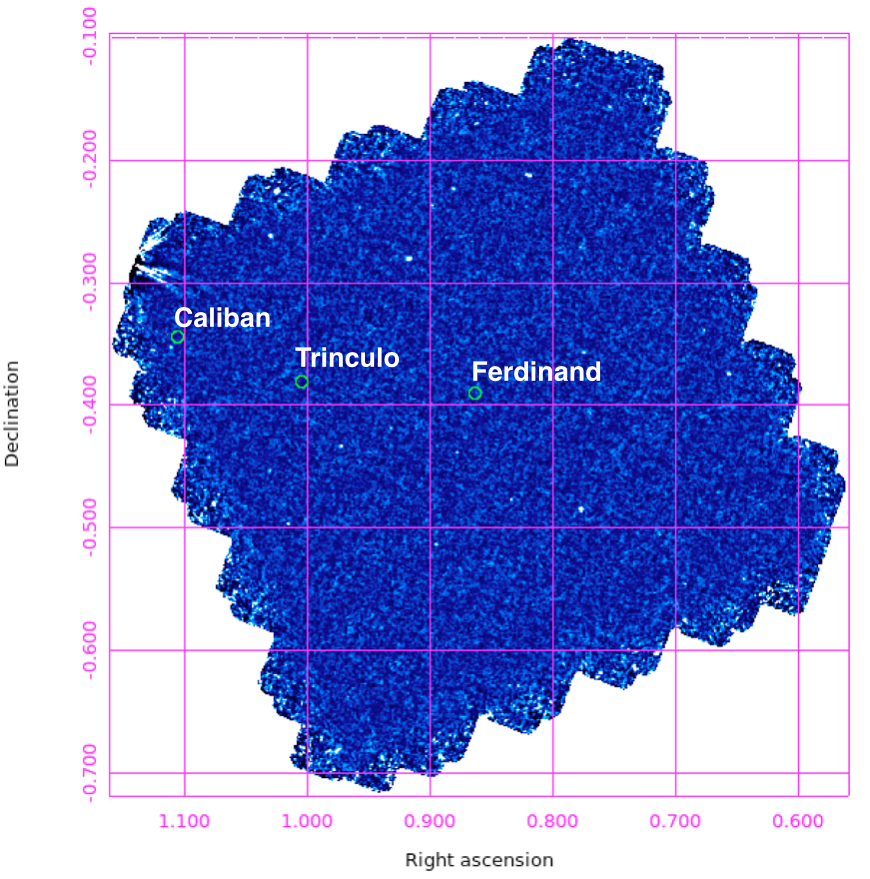}
\caption{Locations of three potentially observable irregular satellites on the Herschel/PACS 70\micron{} Level-2.5 map of two combined OBSIDs 1342237436/37. \label{fig:pacsimagewithmoons}}
\end{figure}
%%%%%%%%%%%%%%

Caliban was identified as potentially present on some far-infrared maps taken with the PACS camera of the Herschel Space Observatory. 
Herschel/PACS observed the environment of Uranus at two epochs: on January 7, 2012 (OBSIDs: 1342236891/92 / scan and cross-scan) and on January 13, 2012 (OBSIDs: 1342237436/37), under the proposal ID~OT1\_ddan01\_1 in both cases. All measurements used the 70/160\,$\mu$m filter combination in all four cases. The data reduction pipeline we used is the same as the one used in the `TNOs are Cool!’ Herschel Open
Time Key Programme \citet{Muller2009}, described in detail in \citet{Kiss2014}, and identical to that we used to reduce the Herschel/PACS maps of Nereid \citep{Kiss2016}. As our aim was to obtain photometry of a point source, we used the {\tt photProject()} task with high-pass filtering to create maps from the time domain detector data. 
The {\tt photProject()} task performs a simple coaddition of the frames using the drizzle method \citep{Fruchter}, and 
the high-pass filtering applies a sliding median-filter on individual pixel timelines. More details on the procedure can be found in The PACS Data Reduction Guide\footnote{see http://herschel.esac.esa.int/hcss-doc-14.0/ for version 14 documentation}.

Maps were created from the detector scans in the co-moving frame of Uranus that was practically identical to that of the satellites due to the small relative velocities of Uranus and the satellites ($<$0\farcs5\,h$^{-1}$).

\begin{figure*}[ht!]
\includegraphics[width=\textwidth]{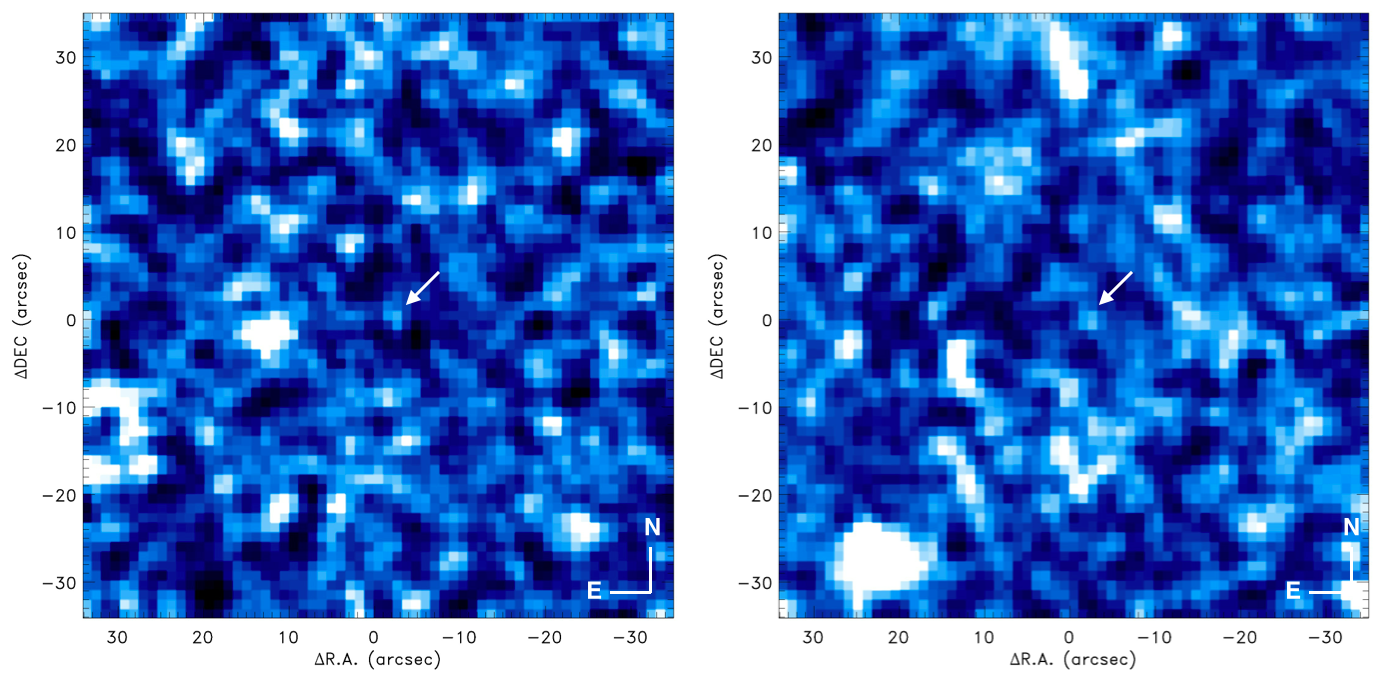}
\caption{Herschel/PACS 70\,$\mu$m images of the region around the expected position of Caliban on 2012 January 7 (left) and January 13 (right). At both epochs a faint source is visible at the expected position (marked by white arrows). At both figures the brightness scaling is set in a way that white colour correspond to fluxes over 3$\sigma$ r.m.s. \label{fig:calibanpacs}}
\end{figure*}

We identified a faint source at both epochs in the 70\,$\mu$m band at the expected location of Caliban, obtained from the NASA Horizons System considering the Herschel-centric observing geometry (see Fig.~\ref{fig:calibanpacs}), and derived a combined flux of F$_{70}$\,=\,1.4$\pm$0.8\,mJy. No obvious source could be identified on the 160\,$\mu$m maps, and the general photometric accuracy obtained using the implanted source method \citep[see][]{Kiss2014} defined a 1-$\sigma$ upper limit of F$_{160}$\,$<$\,3\,mJy for the 160\,$\mu$m brightness of Caliban. 

%%%%%%%%%%%%%%%%%%%%%%%%%%%%%%%%%%%%%%%%%%%%%%%%%%%%%%%%%%%%%%%%%
%%%%%%%%%%%%%%%%%%%%%%%%%%%%%%%%%%%%%%%%%%%%%%%%%%%%%%%%%%%%%%

While Trinculo and Ferdinand could be potentially present on the same set of Herschel/PACS images as Caliban, these were not detected neither at 70, nor at 160\micron, therefore we consider that their flux densities are below 0.8\,mJy and 3\,mJy at 70 and 160\micron, respectively. 

\subsection{Konkoly Observatory 1m-telescope observations of Sycorax}

\begin{figure}[ht!]
\includegraphics[width=8.5cm]{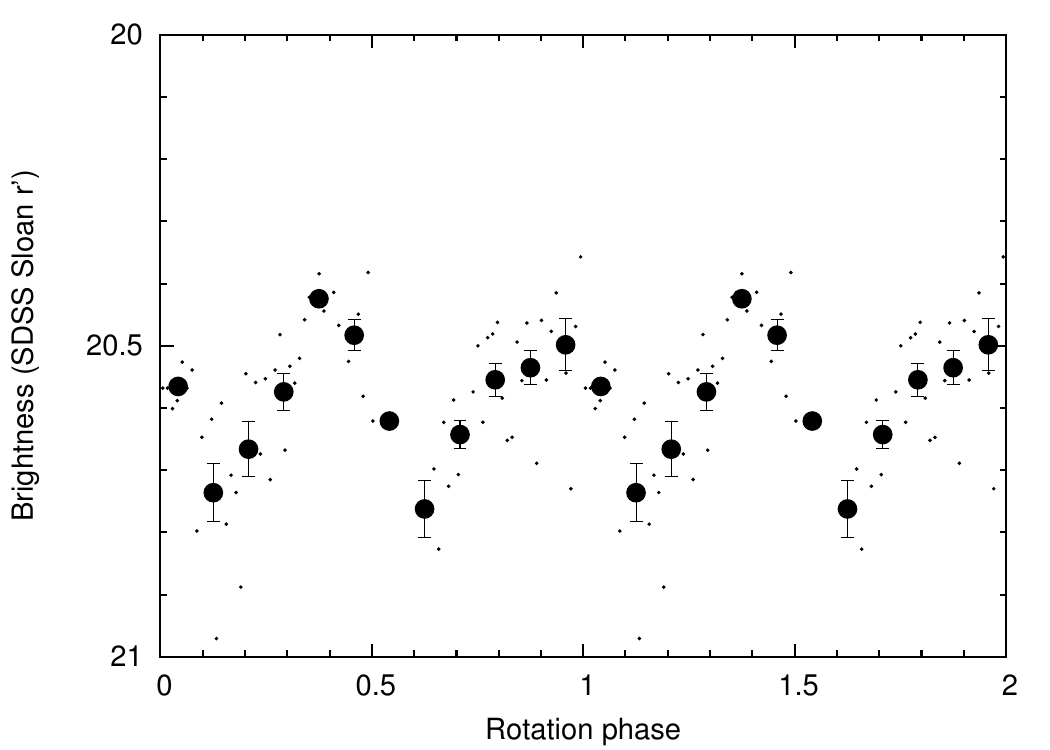}
\caption{Sloan r'-band light curve of Sycorax observed with the 1m-RCC telescope of Konkoly Observatory, folded with the double period of P\,=\,6.9162\,h, obtained from the combination of K2 and 1m-RCC measurements.  \label{fig:sycoraxrcc}}
\end{figure}

Sycorax was also observed on the night of 2015 November 05/06 with the 1-m Ritchey-Cretien Coude (RCC) telescope of the Konkoly Observatory, located at Piszkéstető Mountain Station. In total, 70 frames were acquired with an exposure time of 300 seconds each, using an Andor iXon-888 electron-multiplying CCD (EMCCD) camera. Although it is not so relevant for such long exposures, we note here that the camera was operated in frame transfer readout mode in order to have an effectively zero dead time between the subsequent frames. The frames were taken in Sloan r' filters and used the SDSS-III DR9 catalogue \citep{SDSSDR9} for reference magnitudes of the comparison stars. Standard calibration procedures and aperture photometry were performed using the various tasks of the FITSH package \citep{Pal2012}. The resulted light curve is plotted in Fig.~\ref{fig:sycoraxrcc}.

\section{Thermal emission models\label{sect:thermal}}

To model the thermal emission of some of our targets, infrared monochromatic flux densities were derived from the in-band flux densities applying the appropriate color corrections, based on the surface temperatures of the targets \citep[see ][for the Herschel/PACS and Spitzer/MIPS color corrections, respectively]{Muller2011,mipshandbook}, see also Table~\ref{table:thermal2}. 

We used these monochromatic flux densities and the observing geometry parameters listed in Table~\ref{table:thermal} to constrain the thermal emission models of our targets by calculating the $\chi^2$ values of the modeled and observed monochromatic flux densities \citep[see e.g.][for details]{Vilenius2014}. We applied either the Near-Earth Asteroid Thermal Model \citep[NEATM,][]{Harris98} or a thermophysical model \citep[TPM][]{Lagerros1,Lagerros2,Lagerros3,ML2002} to obtain the model flux densities. TPM was only chosen in the case of Sycorax, as in the case of 
the other satellites with limited amount of thermal data (low degrees of freedom) it is not meaningful to run a complex TPM model over a more simple NEATM one. We used our own NEATM code written in IDL\footnote{Interactive Data Language, Harris Geospatial Solutions}, and a TPM code developed by J.S.V. Lagerros and T.G. M\"uller (see the references above).  

\subsection{Sycorax}\label{sycoraxthermal}
\begin{figure}[ht!]
\includegraphics[width=8.5cm]{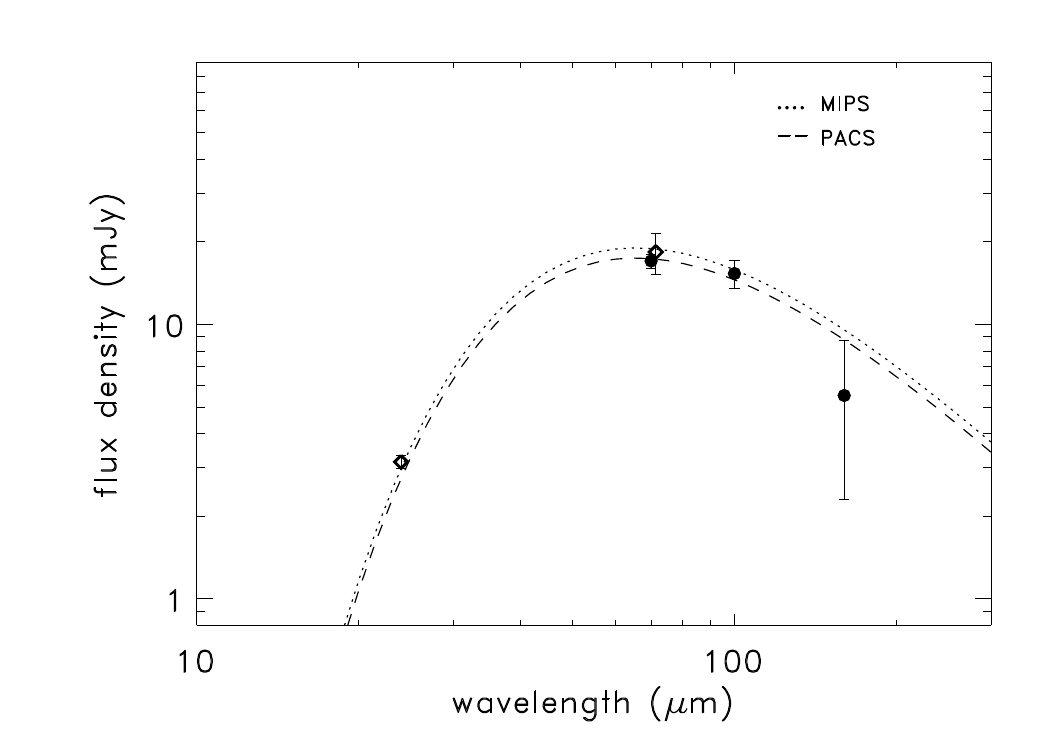}
\caption{Best-fit NEATM results of Sycorax. Flux densities of Spitzer/MIPS (24 and 71\,$\mu$m) and Herschel/PACS (70, 100 and 160\,$\mu$m) measurements are marked by diamonds and filled circles. The dotted and dashed curves correspond to the same set of best-fit NEATM parameters seen at the epoch of the Spitzer/MIPS and Herschel/PACS measurements. \label{fig:sycoraxneatm}}
\end{figure}

The thermal emission of Sycorax was modeled using the flux densities listed in Table~\ref{table:thermal2} applying  a NEATM model. Due to the notable difference in observing geometry at the PACS and MIPS epochs, for a specific model (given size and beaming parameter) the corresponding observation geometries were considered at the PACS and MIPS epochs separately, and the $\chi^2$ values were derived accordingly. The best-fit model is presented in Fig~\ref{fig:sycoraxneatm}, where the flux densities of the MIPS and PACS epochs are shown individually. The NEATM model provided a best-fit effective diameter and albedo estimate of D\,=\,165$\pm$13\,km, 
\geomalb\,=\,0.065$^{+0.015}_{-0.011}$, with a beaming parameter of $\eta$\,=\,1.20$^{+0.25}_{-0.20}$. 

In addition to the NEATM model we also applied thermophysical modeling \citep[TPM, see][and references therein]{ML98,ML2002} considering an absolute magnitude of \absmag\,=\,7\fm50$\pm$0\fm04 \citep{Grav04}, a default slope parameter of G\,=\,0.15 and a wavelength-dependent emissivity \citep{ML2002}. We used P\,=\,6.9162\,h for the rotation period, as determined from the combination of K2 and Konkoly 1m-RCC measurements (see Sect.~\ref{sect:sycoraxrot}); possible thermal inertia values were considered in the $\Gamma$\,=\,0.1--50\,\tiunit{} range and surface roughness values were allowed between $\rho$\,=\,0.1 and 0.9. As the spin-axis orientation of Sycorax is not known, we considered three possible scenarios, a pole-on (spin-axis orientation of $\lambda$\,=\,356\degr, $\beta$\,$\approx$\,0\degr{} in ecliptic coordinates), an equator-on ($\lambda$\,=\,0\degr, $\beta$\,=\,90\degr), and an intermediate one ($\lambda$\,=\,356\degr, $\beta$\,=\,45\degr). The equator-on and intermediate solutions provide a low best-fit reduced $\chi^2$ of \chired\,$<$1, and give very similar best fit values for the size. However, no acceptable solution with sufficiently low \chired{} could be obtained for the pole-on case (\chired\,$>>$1 in all cases) and a non-pole-on configuration is also supported by the presence of a definite visible-range light curve. Our best estimates for the size and albedo are D\,=\,157$_{-15}^{+23}$\,km and \geomalb\,=\,0.07$_{-0.01}^{+0.02}$, associated with a likely intermediate surface roughness ($\rho$\,$\approx$\,0.5) and a close-to-equator-on spin axis configuration. Based on this analysis, very low levels of thermal inertia ($\Gamma$\,$<$1\,\tiunit) can be excluded for Sycorax. 

\begin{figure}[ht!]
\includegraphics[width=8cm]{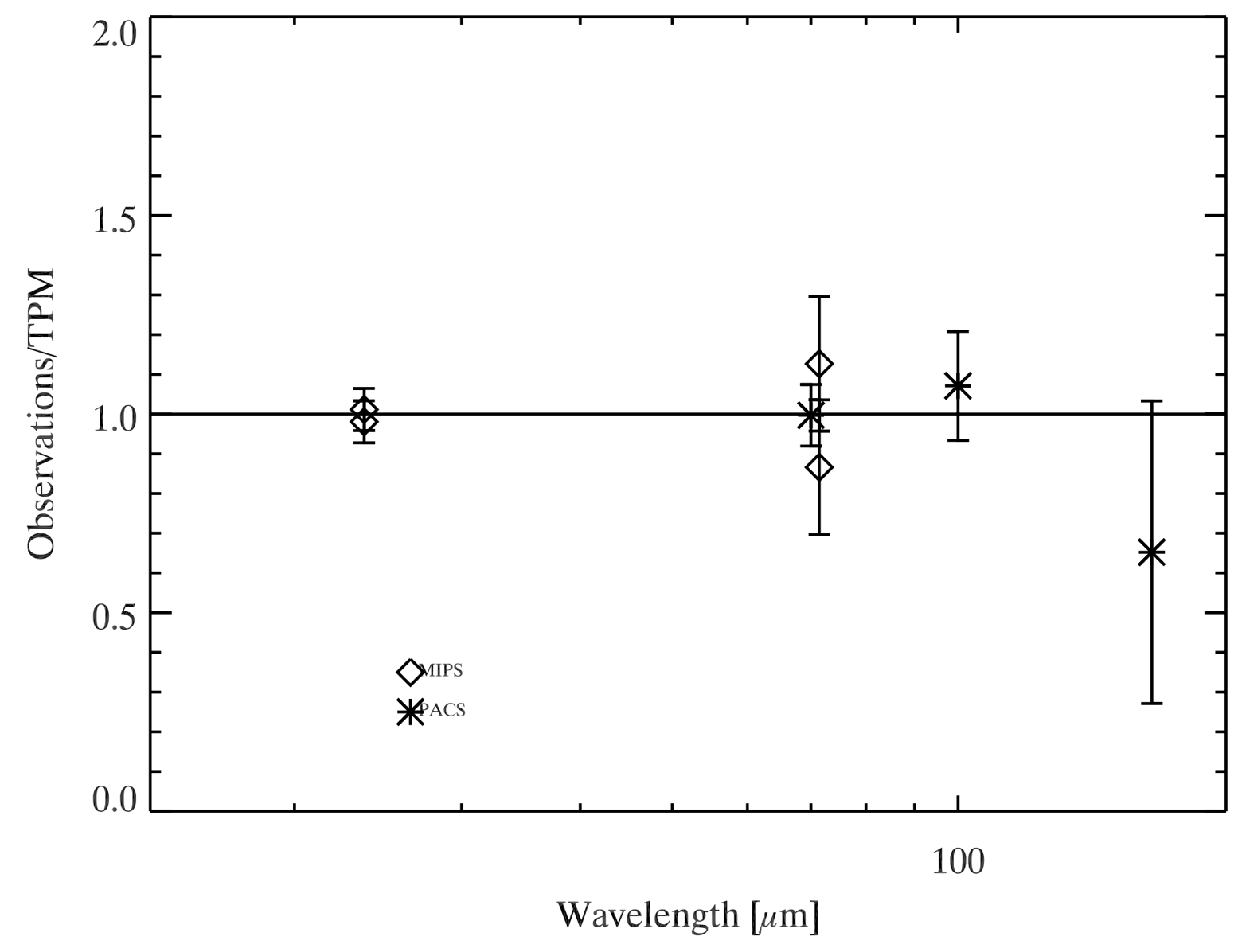}
\caption{Ratio of the observed thermal infrared fluxes of Sycorax to those obtained in the best-fit thermophysical model, as detailed in the text. \label{fig:sycoraxtpm}}
\end{figure}

The diameter and geometric albedo obtained by our analysis is close to the best-fit values obtained by \citet{Lellouch13} using Herschel/PACS data alone (D\,=\,165$_{-42}^{+36}$, \geomalb\,=\,0.049$_{-0.017}^{+0.038}$), but in our case with smaller error bars. However, the beaming parameter is much better constrained with the consideration of the Spitzer/MIPS fluxes, as the Herschel/PACS data could not restrict the models further due to the lack of short wavelength ($\la$40\,$\mu$m) data \citep[$\eta$\,=\,1.26$_{-0.78}^{+0.92}$ in][]{Lellouch13}. Our best-fit $\eta$\,=\,1.20 is very close to the median values obtained for Centaurs and trans-Neptunian objects based on a large sample of Spitzer/MIPS and Herschel/PACS measurements  \citep[$\eta$\,$\approx$\,1.2,][]{stansberry2008,Lellouch13}. The thermal inertia value of $\Gamma$\,=\,3--4\,\tiunit{} we obtained for Sycorax is also in agreement with the $\Gamma$\,=\,5$\pm$1\,\tiunit{} value found by \citet{Lellouch13} for Centaurs at heliocentric distances $<$\,25\,AU. 

\subsection{Small irregular satellites on Herschel/PACS images}

For all potentially detectable satellites we used a NEATM model with a constant emissivity of $\epsilon$\,=\,0.9 to estimate the expected flux densities in the 70\,$\mu$m Herschel/PACS band for a range of diameters/geometric albedos and beaming parameters.  
The beaming parameter $\eta$ was allowed to vary between 0.6 and 1.6, while the diameters were chosen to match a V-band geometric albedo range of 0.01$\leq$\,\geomalb\,$\leq$0.3. Phase integrals were calculated applying both the geometric albedo-dependent phase integral developed for the outer Solar system by \citet{Brucker2009} and the 'standard' value using the canonical slope parameter of G\,=\,0.15 \citep[see e.g.][]{Muinonen2010}. The difference between the 70\,$\mu$m thermal emission flux densities obtained by the two methods were $\leq$4\,$\mu$Jy in all cases, and are negligible for our purposes. 

%The U XVI Caliban was discovered in 1997 by Gladman et al. (1998) \cite{Gladman98}.

%\citet{Maris01} performed a more detailed study of light curve in R band of Caliban using the 3.6 m ESO NTT, La Silla on 1999 October 8 and 9, and determined the BVRI magnitudes. The period of light curve is  $P = 2.6\pm 0.01$\,hr with $A = 0.134 \pm 0.01$\,mag amplitude, furthermore Caliban is redder than the Uranus and regular satellites.

%There is another BVRI color determination of Caliban on 2003 July 27 and 28 by  \citet{Grav04}. This measurement is inconsistent within  $3\sigma$ with \citet{Maris01} in B-V color, they could not explain it.
%Additionally Caliban belongs to the slightly red group based on B-V and V-R colors, furthermore belongs to Caliban family based on its orbit. The absolute magnitude of Caliban was obtained to be H$_{\mathrm V}$\,=\,9\fm16$\pm$0\fm04. 

As described above, Caliban was tentatively detected on Herschel/PACS 70\,$\mu$m maps with a combined monochromatic flux density of F$_{70}$\,=\,1.4$\pm$0.8\,mJy. A generally assumed dark surface of $p_\mathrm{V}$\,$\approx$\,0.04 and the corresponding diameter of $\sim$70\,km would produce a 70\,$\mu$m flux density of F$_{70}$\,$>$\,2.4\,mJy, easily detectable on our Herschel/PACS maps, over 3$\sigma$ of the 0.8\,mJy 70\,$\mu$m flux uncertainty of the Herschel/PACS maps. The fact that the detected flux density of Caliban is \emph{smaller} than that already indicates a brighter surface. 
We obtained D\,=\,42$_{-12}^{+20}$\,km and \geomalb\,=\,0.22$_{-0.12}^{+0.20}$ as the best-fit NEATM solution using the single available 70\,$\mu$m flux density, with no real constraint on the beaming parameter in our originally chosen 0.6\,$\le$\,$\eta$\,$\le$\,1.6 range, with our best-fit model having a corresponding $\eta$ of 0.8. The 70\,$\mu$m flux of Caliban can, however, be equally well fitted with D\,$\approx$50\,km and \geomalb\,$\approx$\,0.15, using a higher $\eta$ value of $\sim$1.4. The uncertainties in the size and albedo due to the unconstrained beaming parameter are reflected in the errors of the best-fit values quoted above. 
The geometric albedo we obtained for Caliban may seem to be surprisingly high in the Uranian irregular satellite system, taking into account the low albedo of Sycorax. On the other hand, relatively bright surfaces exist among other irregular satellites, e.g. Nereid has a surface with \geomalb\,$>$\,20\% \citep[][]{Kiss2016}, but it is notably larger than Caliban, $\sim$350\,km in effective diameter.  

\begin{figure}[ht!]
\includegraphics[width=8.5cm]{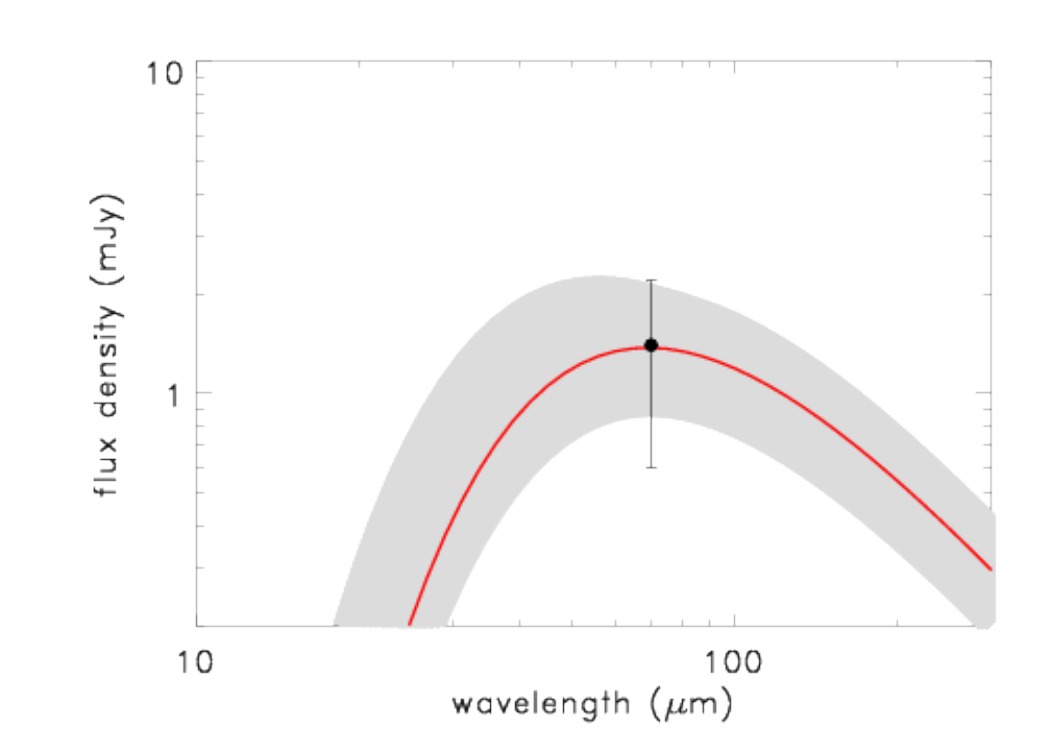}
\caption{NEATM modeling results for Caliban. The black filled circle with error bar shows the only available measurement at 70\,$\mu$m (Herschel/PACS). The red curve represent the best-fit result (D\,=\,42\,km, \geomalb\,=\,0.22 and $\eta$\,=\,0.8), the gray area contains the model curves that are compatible with the 70\,$\mu$m flux (see the text for details). \label{fig:calibanneatm}}
\end{figure}
%%%%%
%\begin{table*}
%\begin{tabular}{ccccccc}
%\hline
%Satellite 	& H$_{V/R}$ & V--R & B--V & S$'$& Ref.\\
%          	& (mag)     & (mag)& (mag)& (\%/1000\,\AA)&\\
%\hline          
%Sycorax   	& 7.50$\pm$0.04(G04) 	& 0.62$\pm$0.01 & 0.78$\pm$0.02	& 27.33$\pm$1.045 	& G04\\
%\hline
%Caliban   	& 9.16$\pm$0.00()		& 0.57$\pm$0.03	& 0.84$\pm$0.03	& 22.1$\pm$3.13 	& G04\\
%\hline
%Trinculo   	& 11.9$\pm$0.2()		& 0.35$\pm$0.19	& 1.09$\pm$0.4	& -1.1$\pm$20	 	& G04\\
%\hline
%Prospero   	& 10.6$\pm$0.1()		& 0.39$\pm$0.04	& 0.8$\pm$0.06	& 3.16$\pm$4.22	 	& G04\\
%\hline
%Setebos   	& 10.6$\pm$0.1()		& 0.35$\pm$0.03	& 0.77$\pm$0.06	& -1.1$\pm$3.16	 	& G04\\
%\hline
%Francisco  	& 		& 	& 	& 	 	& \\
%\hline
%Ferdinand  	& 		& 	& 	& 	 	& \\
%\hline
	%
%\end{tabular}
%\caption{Summary of previous and present observations of the Uranian irregular satellites Sycorax, Caliban, Setebos, Ferdinand and Prospero. The references used in the table are: M01 -- \citet{Maris01};\label{table:colors}}
%\end{table*}
%%%%%%%%%%%%%%%%%%%%%%%%%%%%%%%%%%%%%%%%%

\begin{table*}
\footnotesize
\begin{tabular}{cccc|ccccc}
\hline
            & \multicolumn{3}{c}{Previous works} & \multicolumn{3}{|c}{This work} \\
\hline            
Satellite 	& P & $\Delta$m &Ref. & f$_0$ & P & $\Delta$m & comm. & P$_{s}$\\
          	& (h)     & (mag)& & (cycle\,day$^{-1}$) & (h) & (mag) & & (h)\\
\hline          
Sycorax   	& 4.12$\pm$0.04 & 0.032$\pm$0.008$^s$ &	M01 & 6.9374$\pm$0.0083 & 6.9190$\pm$0.0082& 0.121$\pm$0.020 & K2,d & 67.3423 \\    
            & 3.60$\pm$0.02 & 0.067$\pm$0.004$^s$ & M07 & 6.9402$\pm$0.0013 & 6.9162$\pm$0.0013& 0.120$\pm$0.019 & K2+1m,d \\
\hline
Caliban 	& 2.66$\pm$0.04	& 0.13$\pm$0.01$^s$   & M01 & 4.8249$\pm$0.0092 & 9.948$\pm$0.019 & 0.16$\pm$0.03 & K2,d & 7.0026 \\
\hline
%Trinculo 	& && \\
%\hline
Prospero 	& 4.55$\pm$0.04 & 0.22$\pm$0.03$^s$   & M07 & 3.359$\pm$0.044 & 7.145$\pm$0.092 & 0.41$\pm$0.07 & K2,s & 16.5871 \\
\hline
Setebos 	& 4.38$\pm$0.05	& 0.189$\pm$0.03$^s$  & M07 & 5.640$\pm$0.022 & 4.255$\pm$0.017 & 0.27$\pm$0.06 & K2,s & 13.1705 \\
\hline
%Francisco 	& & & \\
%\hline
Ferdinand 	& & & & 2.027$\pm$0.039 & 11.84$\pm$0.22 & 0.54$\pm$0.09 & K2,s & 82.715\\
\hline
%%%
\end{tabular}
\caption{Summary of previous and present light curve observations of the Uranian irregular satellites Sycorax, Caliban, Prospero, Setebos and Ferdinand with the most likely rotation periods identified. 
The $\Delta m$ light curve amplitudes are peak-to-peak values, unless marked with an 's' upper index -- in this case they represent a sinusoidal amplitude.   
The references used in the table are: M01 -- \citet{Maris01}; M07 -- \citet{Maris07}. 
In the last column 'K2' and '1m' indicates data obtained from the K2 mission and with 1m-telescope of Konkoly Observatory, while 'd' and 's' mark single-peak and double-peak rotation periods. The last column (P$_s$) lists the stroboscopic periods calculated from the single peak periods.  
\label{table:periods}}
\end{table*}

Trinculo and Ferdinand were not detected on the Herschel/PACS images. However, considering the 0.8\,mJy 70\,$\mu$m flux uncertainties as an upper limit for both targets we can put some constraints on their geometric albedos and diameters. We note that for Ferdinand we calculated the R-band absolute magnitude using data in the Minor Planet Circular MPEC-2003-S105, assuming an R-band specific linear phase correction of $\beta_{\mathrm R}$\,=\,0.119 \citep{Belskaya2008} and obtained H$_\mathrm{R}$\,=\,12.06$\pm$0.15\,mag. For Trinculo we used the value provided by \citet{Grav04} (see also Table~\ref{table:thermal2}). The 0.8\,mJy 1$\sigma$ upper limit indicates a geometric albedo of \geomalb\,$>$0.03 for both satellites, and correspondingly their diameters are D\,$<$\,50\,km. 

%%%%%%%%%%%%%%%%%%%%%%%%%%%%%%%%%%%%%%%%%
\section{Rotational characteristics from K2 measurements \label{sect:rotation}}

We searched for significant periodicities using the Fourier method as implemented in the \emph{Period04} program package \citep{Lenz} and also the Lomb-Scargle periodogram in the \emph{gatspy} Python package\footnote{https://github.com/astroML/gatspy/}. 
We got very similar results in several test cases, therefore we decided to stick to the Lomb-Scargle periods. We note that the errors of the individual photometric points are taken into account. Only those signals were considered that were significant on the 3$\sigma$-level compared to the background local noise periodogram. We phase-folded the light curves with the best period and its double value, then decided which gave a better fit based on a visual inspection. As we have shown previously \citep{Szabo2016} period determination of Solar system object with K2 long cadence measurements is solid if the coverage exceeds five days and the duty cycle is above 60\%. These conditions are fulfilled for three of our targets, as we have shown in Table~\ref{tab:keplerlog} already. The other two, Setebos and Ferdinand, have lower duty cycles, but the long baseline of the observations compensates for them. Overall, we were able to derive reliable solutions for most Uranian irregular satellites in our sample. The results are presented in Figs.~\ref{fig:rot1}, \ref{fig:rot2} and in Table~\ref{table:periods} below. Table~\ref{table:periods} also lists the rotation periods obtained from previous investigations. We emphasize that we do not accept all formally significant periods, but simply choose the one with the highest amplitude.

\subsection{Sycorax \label{sect:sycoraxrot}} 

\citet{Maris01} performed the first detailed study of the light curve in the $R$ band using the 3.6\,m ESO NTT telescope at La Silla, on 1999 October 8 and 9. The amplitude of light variation they found was $A$\,=\,0\fm032$\pm$0\fm008 with a P\,=\,4.12$\pm$0.04\,h period. Measurements taken with the VLT in 2005 \citep{Maris07} provided the most likely light curve period and amplitude of $P$\,=\,3.6$\pm$0.02\,h and $A$\,=\,0\fm067$\pm$0\fm004. 

\begin{figure*}
\includegraphics[width=\textwidth]{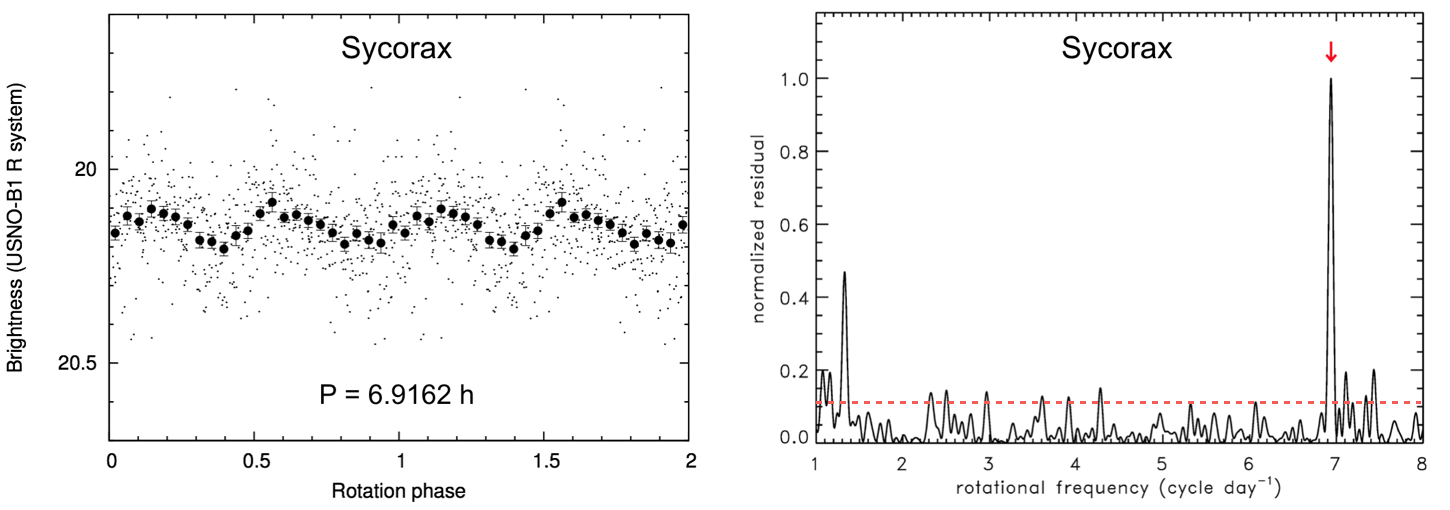}
\includegraphics[width=\textwidth]{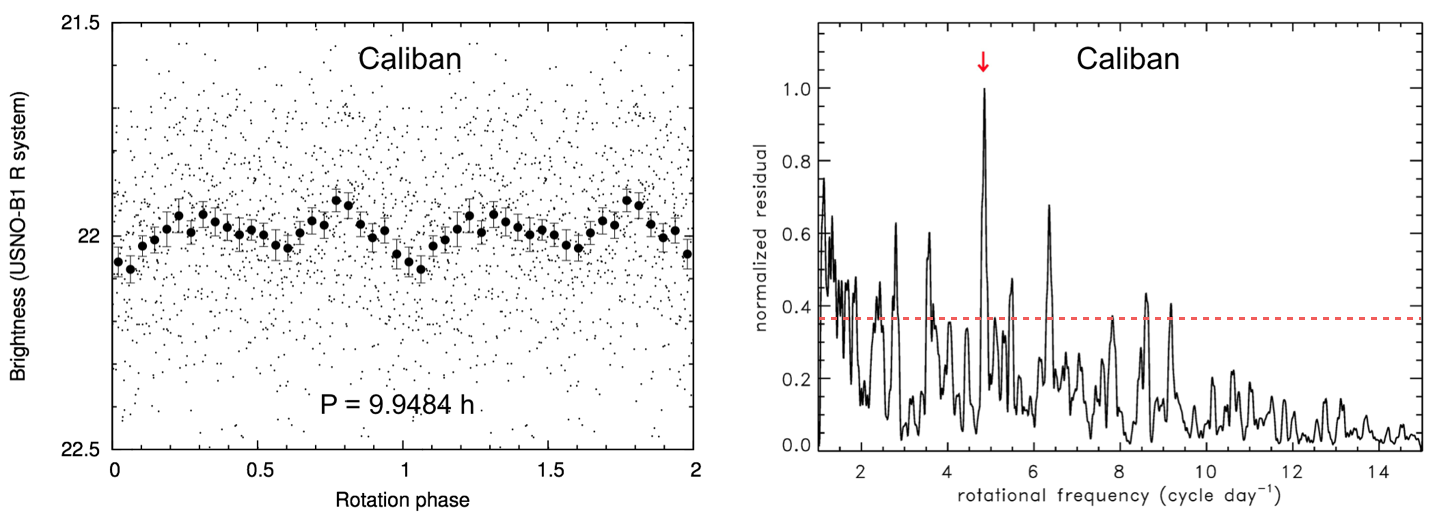}
\includegraphics[width=\textwidth]{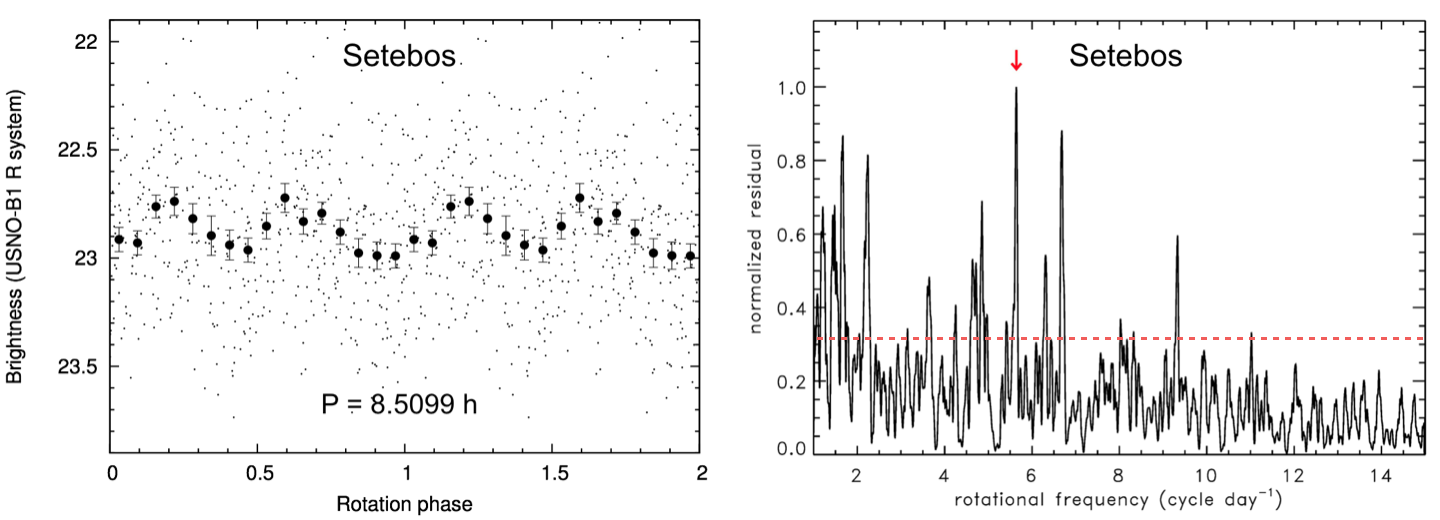}
\caption{Folded K2 (USNO-B1 R-band) light curves and normalized residual vs.\ frequency diagrams of Sycorax, Caliban and Setebos. The folding periods are indicated on the left panels. The most likely frequencies are marked by red arrows in the normalized residual figures. The 3\,$\sigma$ significance levels are indicated by red horizontal lines in the normalized residual vs.\ frequency diagrams. 
\label{fig:rot1}}
\end{figure*}

Our K2 measurements revealed a well-defined rotational period with a frequency of f\,=\,6.9374$\pm$0.0083\,cycle\,day$^{-1}$ (P\,=\,3.458$\pm$0.001\,h, see Fig.~\ref{fig:rot1}). We assume that the light curve of Sycorax is double-peaked which is supported by the slight asymmetry of the light curve when it is folded with the half frequency / double rotation period (Fig.~\ref{fig:rot1}). This gives a double-peaked rotation period of P\,=\,6.9190$\pm$0.0082\,h. The single-peaked period is very close to that obtained by \citet{Maris07} using VLT measurements. The peak-to-peak light curve amplitude obtained from K2 data is A\,=\,0\fm12$\pm$0\fm02, consistent with that obtained by \citet{Maris07} as sinusoidal amplitude.  

There is a second significant peak on the frequency diagram at 0.35\,cycle\,day$^{-1}$ (Fig. 9, upmost panel). Such secondary periods are often explained by tumbling rotation or a companion. Although a companion around Sycorax (a moon of a moon) would be an intriguing possibility, we suggest that this second peak is a sampling artifact. The interaction of a strictly periodic sampling (such as that of \textit{Kepler}) and a strictly periodic process (such as a rotation of the asteroid) results in periodic phase shifts in the sampling, and hence, an emergence of a stroboscopic period that can modulate timing, brightness, shape modulations etc. In \citet{Szabo2013} we calculated the stroboscopic period as $P_s /P = 1/ {\rm min}( \, ] P/C[ , \, ] 1-P/C [ \, )$, where $P_s$ is the stroboscopic period, $P$ is the observed actual period, C is the cadence, and $] \, [$ denotes the fractional part. Substituting the rotation period (6.9162$\pm$0.0013\,h) and the cadence (1765.5\,s) into the formula, the stroboscopic period is calculated to be 9.5--10.0-times the rotation period. Since the ratio of the two detected periods is 9.97, this is perfectly compatible with the stroboscopic origin of the long period brightness variation. Stroboscopic periods were calculated from the single peak periods of the other targets, too (see below), as listed in Table~\ref{table:periods}. We note that a stroboscopic period is not accepted simply by an amplitude criterion as a possible rotation period, but should be rejected due to its stroboscopic nature.  

Since the series of ground-based observations with the 1m telescope of Konkoly Observatory was roughly 3 months before the K2 measurements and the precision of K2 measurements are rather fine, one can extrapolate the rotation cycle numbers in an unambiguous manner for such a relatively short period. This allows us to combine the two series of measurements (K2 and 1m RCC) to get an accurate rotational frequency, for which we obtained n=3.47012$\pm$0.00067\,cycle\,day$^{-1}$, i.e. P=6.9162$\pm$0.0013\,h. 

%%%%%%%%%%%%%%%%%%%
%%%%%%%%%%%%%%%%%%%
%%%%%%%%%%%%%%%%%%%
\begin{figure*}
\includegraphics[width=\textwidth]{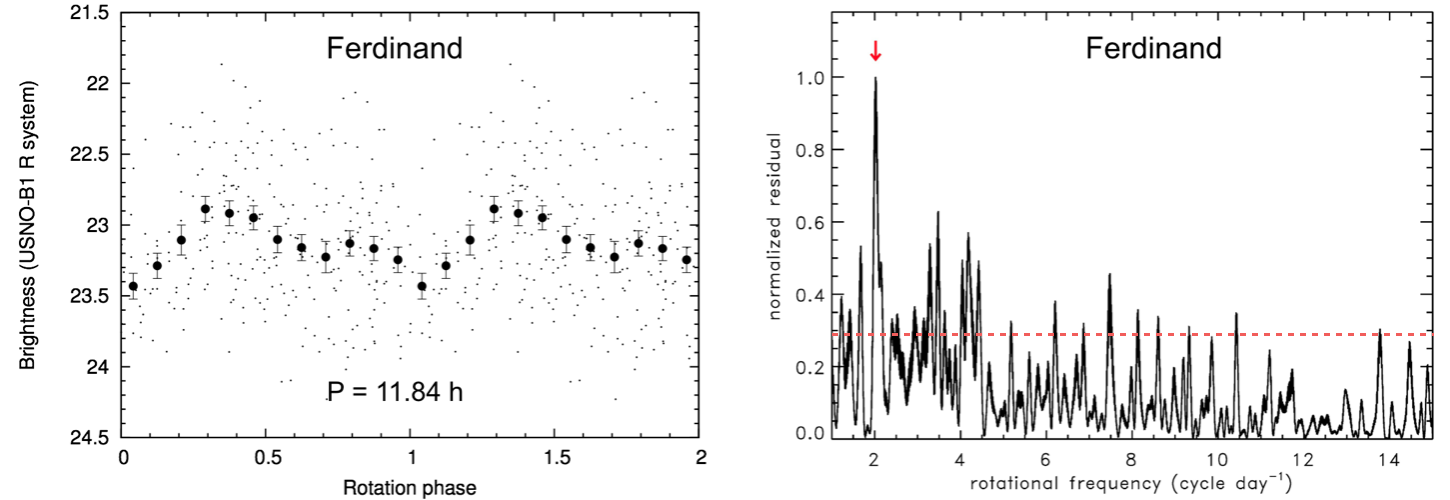}
\includegraphics[width=\textwidth]{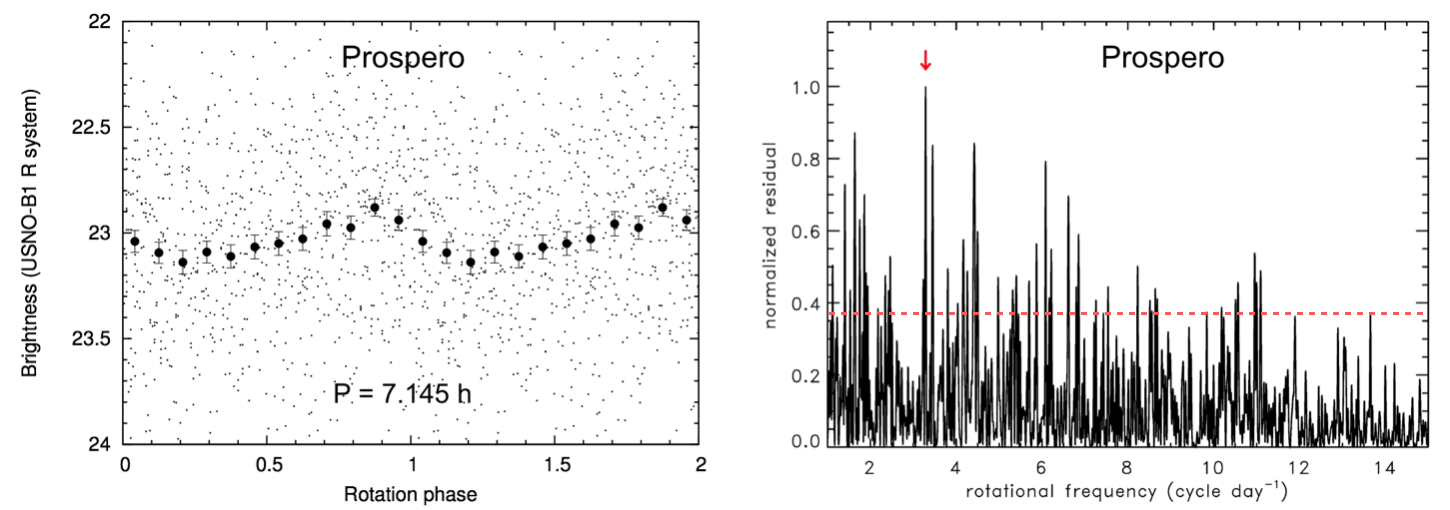}
\caption{The same as Fig.~\ref{fig:rot1} but for Ferdinand and Prospero. \label{fig:rot2}}
\end{figure*}
%%%%%%%%%%%%%%%%%%%

\subsection{Smaller satellites} 

\paragraph{Caliban:} \citet{Maris01} determined a light curve period of 2.66\,h that is not confirmed by our data (see Fig.~\ref{fig:rot1}). Instead, we obtained a most likely frequency of f\,=\,4.8249$\pm$0.0092\,cycle\,day$^{-1}$ and the asymmetry of the folded light curve indicates that the real rotation period corresponds to the half frequency, i.e. P\,=\,4.9742$\pm$0.0095\,h. The single peak period of Caliban is the only one apart from that of Sycorax for which the corresponding stroboscopic frequency (P$_s$\,=\,7.0026\,h or f$_s$\,=\,3.4256\,cycle\,day$^{-1}$) is close to a significant peak in Fig.~\ref{fig:rot1}. 
%A peak with a similar significance can be identified at $\sim$6.5\,h, the stroboscopic period of the double rotation period identified.

\paragraph{Prospero:} In our analysis the least unambiguous light curve period was obtained for Prospero. We identified the most likely period of 3.359$\pm$0.044\,cycle\,day$^{-1}$ (P\,=\,7.145$\pm$0.092\,h), but a strong, secondary peak is also visible at f\,=\,4.415$\pm$0.045\,cycle\,day$^{-1}$ that corresponds to a single-peak rotation period of P\,=\,5.346$\pm$0.055\,h, very close to the light curve period obtained by \citet{Maris07}. 

\paragraph{Setebos:} For this satellite we confirm the light curve period of 4.38$\pm$0.05\,h obtained by \citet{Maris07} as we derived a most likely rotation period of P\,=\,4.255$\pm$0.017\,h, very close to the previously mentioned value, without indication of a double-peak light curve.  

\paragraph{Ferdinand:} The light curve period of Ferdinand was not determined earlier.  The most likely frequency of 2.027$\pm$0.039\,cycle day$^{-1}$ corresponds to a rather long rotation period of 11.84$\pm$0.22\,h, the longest one in our sample (assuming a single-peak light curve). Such long rotation periods are, however, not rare and present e.g. in the Saturnian system where the rotation periods of 10 from 16 irregular satellites in a recently studied sample show rotation periods longer than that of Ferdinand \citep{Denk2013}. 

%%%%%%%%%%%%%%%%%%%%%%%%%%%%%%%%%%%%%%%%%%%%%%%%%%%%%%%%%%%%%%%%%%%%%
\section{Comparison with the rotational characteristics of other irregular satellites and asteroids\label{sect:rotcomp}}

\begin{figure*}[ht!]
\includegraphics[width=\textwidth]{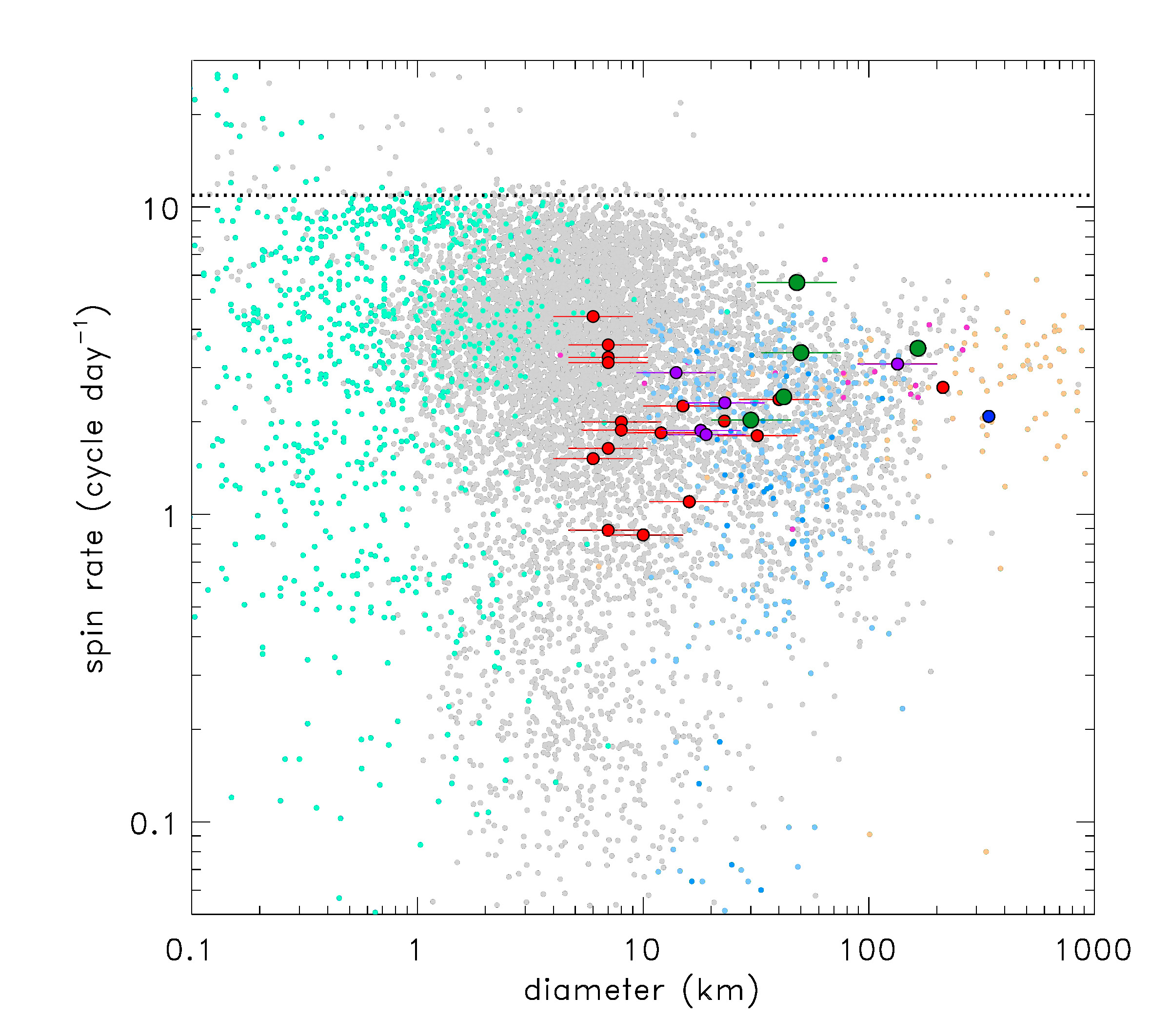}
\caption{Diameter vs.\ rotational frequency of irregular satellites and asteroids. Filled green circles mark the four Uranian irregulars discussed in this paper. Purple and red filled circles mark the Jovian \citep{Luu,Pilcher} and Saturnian irregular satellites \citep{Denk2013,Bauer}, while the blue filled circle mark the Neptunian irregular satellite Nereid \citep{Kiss2016}. Horizontal bars around some satellites indicates the possible size range due to a large uncertainty in the estimation of the effective diameter from the visible range absolute magnitude. Pale-colored dots in the background correspond to the data of various minor planet populations throughout the Solar system (green: near-Earth asteroids; gray: main belt asteroids; blue: Jovian Trojans; magenta: Centaurs; orange: trans-Neptunian objects).  \label{fig:spin}}
\end{figure*}

Rotation of small body populations in the Solar system is often characterized by the so-called spin barrier, a  critical  rotation  period  at  which  a  rubble  pile  asteroid  would  fly  apart  due  to  its  centripetal  acceleration. This  spin  barrier  is  well established for Main Belt asteroids, the critical rotation period is $\sim$2.2 h (dashed horizontal line in Fig.~\ref{fig:spin}), resulting in a $\rho_{crit}$ critical density estimate of $\sim$2.0\,g\,cm$^{-3}$, using the formula by \citet{Pravec+Harris}. In Fig.~\ref{fig:spin} we plot the rotation period versus size for various small body populations as well as for irregular satellites of the giant planet systems.

%%%%%%%%%%%%%%%%%%%
%%%%%%%%%%%%%%%%%%%
%

Considering our sample, the \emph{median} rotational frequencies are notably higher in the Uranian system ($\approx$3.4\,cycle\,day$^{-1}$) than in the other giant planet systems ($\approx$2\,cycle\,day$^{-1}$ for Jupiter, Saturn and Neptune). Also, assuming the double-peak rotation periods for Sycorax and Caliban provides us with critical densities $\rho_\mathrm{crit}$\,$\le$\,0.76\,g\,cm$^{-3}$. When the single-peak periods are considered for all Uranian irregular satellites $\rho_\mathrm{crit}$ still remains below 1\,g\,cm$^{-3}$. These values are below the upper limits of $\sim$2\,g\,cm$^{-3}$ of main belt asteroids, but higher than the $\sim$0.5\,g\,cm$^{-3}$ obtained for e.g. Jovian Trojans \citep{Szabo2017}, the typical densities of cometary nuclei and trans-Neptunian objects \citep{AHearn2011,Brown2013,Vilenius2014}, and also those critical densities that can be estimated for the irregular satellites of the other giant planet systems, based on rotational light curves alone. However, e.g., the mass of the largest Jovian irregular Himalia has been estimated from its perturbations on other satellites \citep{Emelyanov}, and it gives an independent estimate on the density, $\rho$\,$>$\,2.6\,g\,cm$^{-3}$, using the size obtained during the Cassini flyby \citep[effective radius of $\sim$67\,km][]{Porco03}. This is significantly larger than the critical density of $\sim$0.2\,g\,cm$^{-3}$ that can be obtained from the rotation period of 7.78\,h and the light curve amplitude of 0\fm20$\pm$0\fm01 \citep{Pilcher}. 

The Uranian irregular satellites in our sample are in the size range where Maxwellian distribution of rotational frequencies starts for main belt asteroids \citep[D\,$\ga$\,40\,km,][]{Pravec2002}. While our sample is small and certainly not unbiased, the large median rotational frequency of the Uranian irregular satellites (3.4\,cycle\,day$^{-1}$) may indicate that this irregular satellite system had a collisional evolution different from those around Jupiter and Saturn and Uranian irregulars suffered from a higher number and/or more energetic collisions. The median rotation period of 7.1\,h in the Uranian system is close to that obtained for Centaurs \citep[7.35\,h,][]{Duffard2009} and somewhat smaller than that of trans-Neptunian objects \citep[8.6\,h,][]{Thirouin2014}, however, these populations certainly went through a different collisional evolution than that of the Uranian irregular satellite system. 

\section{The albedo-colour diversity of irregular satellites \label{sect:albedocolor}}

%%%%%%%%%%%%%%%%%%%
\begin{figure}[ht!]
\includegraphics[width=8.5cm]{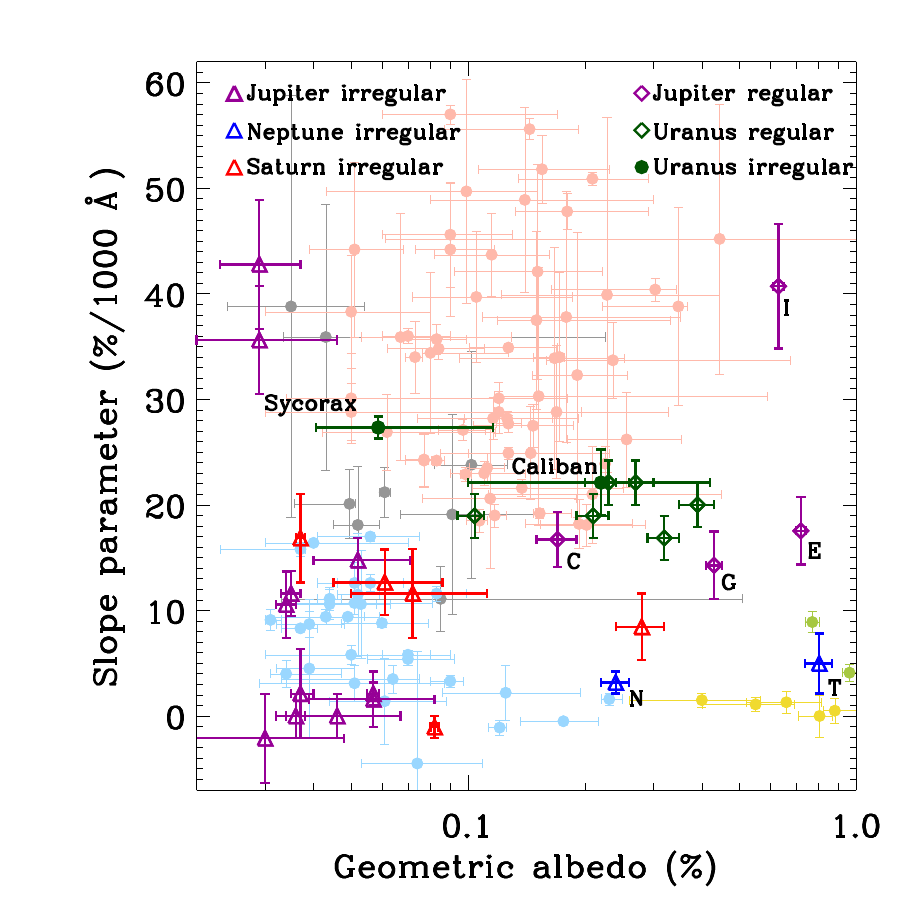}
\caption{Albedo and colour characteristics of irregular satellites of the giant planet systems. 
Purple, red, green and blue correspond to the satellites of Jupiter, Saturn, Uranus and Neptune, respectively. Error bars mark the uncertainties in geometric albedo and color. 
Pale dots in the background correspond to the data of outer Solar system objects (Centaurs and trans-Neptunian objects), taken from \citet{L14}. Europa, Io, Ganymede, Callisto, Nereid and Triton are marked with their initials.
References: \cite{Buratti1983}, \cite{Cruikshank1982}, \cite{Denk2013},   \citet{Grav03,Grav04,Grav2015}, \cite{Hick2004}, \cite{Karkoschka2001}, \cite{Kiss2016}, \cite{Lellouch13}, \cite{Millis1975}, \cite{Morrison2009},  \cite{Rettig2001}, \cite{Showalter2006}, \cite{Simonelli1984}, \cite{Smith1989}.
\label{fig:pvcolour}}
\end{figure}
%%%%%%%%%%%%%%%%%%%

In Fig.~\ref{fig:pvcolour} we plotted the colors \citep[represented by spectral slopes,][]{LJ} versus the geometric albedos of those irregular satellites for which these information were available. 
The Jovian irregular satellites are typically found in the same albedo-color region as the Centaurs/trans-Neptunian objects with dark-neutral surfaces (pale blue dots in Fig.~\ref{fig:pvcolour}). This is also the characteristic region for cometary nuclei and Jovian Trojan asteroids \citep[see e.g.][]{L14}. Only two Jovian irregular satellites show red surfaces, both extremely red and dark, already outside the bright-red group of outer Solar systems objects (pale red dots in Fig.~\ref{fig:pvcolour}). Saturnian irregulars (red symbols) are obviously in the dark-neutral group. Here we included Hyperion, too (highest albedo Saturnian point in Fig.~\ref{fig:pvcolour}), although this is strictly speaking not an irregular satellite, but shows characteristics different from the typical regular Saturnian satellites (elongated shape, highly cratered surface, likely porous interior). The Neptunian irregular, Nereid is also a likely dark-neutral object. Triton, however, is clearly distinct from all other irregulars, and, as it is expected due to its large size, resemble more the group of large dwarf planets than any other irregular satellites -- in that case internal processes (cryovolcanism) may have significantly altered the original surface. The surface of Triton, as well as those of the large regular satellites are more similar to the largest dwarf planets (green symbols in Fig.~\ref{fig:pvcolour}) and the members of the Haumea collisional family (yellow symbols).   

The two irregular Uranian satellites, Sycorax and Caliban, for which albedo and color data are available both seem to fall into the bright-red group, along with some regular Uranian satellites (Puck, Miranda, Ariel, Umbriel, Titania, Oberon). Currently no other irregular satellites in other giant planet system can be assigned to this albedo-color group. 

Although our sample is limited, the location of the irregular satellites on the albedo-color diagram may indicate that the surfaces of satellites in the Uranian system may resemble those of the bright-red trans-Neptunian objects. Irregular satellites in the Jovian and Saturnian systems, and also Nereid are generally darker and more neutral in color. If the surfaces of the Uranian irregular satellites and those in other giant planet systems are intrinsically different, and not a consequence of a different evolution of surfaces, this may be a further indication of a compositional discontinuity in the young Solar system. This discontinuity should have existed close to the heliocentric distance of Uranus, caused by the same processes that induced the bimodality among Centaurs and trans-Neptunian objects \citep{L14}. 

%%%%%%%%%%%%%%%%%%%
\begin{acknowledgements}
The research leading to these results has received funding from the European Unions Horizon 2020 Research and Innovation Programme, under Grant Agreement No. 687378; from the K-115709, PD-116175, and GINOP-2.3.2-15-2016-00003 grants of the National Research, Development and Innovation Office (NKFIH, Hungary); and from the LP2012-31 grant of the Hungarian Academy of Sciences.  L.~M. was supported by the J\'anos Bolyai Research Scholarship of the Hungarian Academy of Sciences. Funding for the \textit{Kepler} and K2 missions are provided by the NASA Science Mission Directorate. The data presented in this paper were obtained from the Mikulski Archive for Space Telescopes (MAST). STScI is operated by the Association of Universities for Research in Astronomy, Inc., under NASA contract NAS5-26555. Support for MAST for non-HST data is provided by the NASA Office of Space Science via grant NNX09AF08G and by other grants and contracts. This work is based in part on archival data obtained with the Spitzer Space Telescope, which is operated by the Jet Propulsion Laboratory, California Institute of Technology under a contract with NASA. The authors thank the hospitality the Veszpr\'em Regional Centre of the Hungarian Academy of Sciences (MTA VEAB), where part of this project was carried out. We also thank our referee for the helpful comments and suggestions. 
\end{acknowledgements}

\software{FITSH \citep{Pal2012}, Period04 \citep{Lenz}, gatspy (https://github.com/astroML/gatspy/)}
%%%%%%%%%%%%%%%%%%%%%%%%%%%%%%%%%%


\begin{thebibliography}{}

%%%%
\bibitem[A'Hearn(2011)]{AHearn2011}
A’Hearn, M. F. 2011, ARA\&A, 49, 281
%%%%
\bibitem[Ahn et al.(2012)]{SDSSDR9}
Ahn, C.P., Alexandroff, R., Allende, P.C., et al., 2012, ApJS, 203, 21
%%%%
%\bibitem[Aknes \& Grav(2005)]{Aknes05}
%Aknes, K. \& Grav, T., A\&A 441, 815 
% Orbit correction without variational equations The orbits of Caliban, 46P/Wirtanen and 67P/Churyumov-Gerasimenko
%%
\bibitem[Bauer et al.(2004)]{Bauer}
Bauer, J.M., Buratti, B.J., Simonelli, D.P., Owen, W.M., 2004, ApJL, 610, L57
%%
\bibitem[Belskaya et al.(2008)]{Belskaya2008}
Belskaya, I.N., Levasseur-Regourd, A.-C., Shkuratov, Y.G., Muinonen, K., 2008, in The Solar System Beyond Neptune, Surface Properties of Kuiper Belt Objects and Centaurs from Photometry and Polarimetry, p.115 (Tucson, AZ: Univ. Arizona Press)  
%%
\bibitem[Brown(2013)]{Brown2013}
Brown, M. E., 2013, ApJ, 778, L34
%%
\bibitem[Buratti \& Veverka(1983)]{Buratti1983}
Bonniel Buratti \& Joseph Veverka, 1983, Icarus, 55, 93
%Voyager photometry os Europa
%%
\bibitem[Brucker et al.(2009)]{Brucker2009}
Brucker, M., Grundy, W.M., Stansberry, J.A., et al., 2009, Icarus, 201, 284
%%
\bibitem[Colbert et al.(2011)]{mipshandbook}
Colbert, J., and the MIPS Instrument and MIPS Instrument Support Teams, MIPS Instrument Handbook, 2011, Version 3.0 ()
%%
%\bibitem[Colombo \& Franklin(1971)]{Colombo}
%Colombo, G., \& Franklin, F. A. 1971, Icarus, 15, 186 
%On the formation of the outer satellite groups of Jupiter
%%
\bibitem[Cruikshank \& Brown(1982)]{Cruikshank1982}
Cruikshank, D.P. \& Brown, R.H., 1982, Icarus, 50, 82
%Surface composition and radius of Hyperion
%%
\bibitem[Denk \& Mottola(2013)]{Denk2013}	
Denk, T. \& Mottola, S., 2013, American Astronomical Society, DPS meeting \#45, id.406.08 
%%
\bibitem[Denk \& Mottola(2014)]{Denk2014}	
Denk, T. \& Mottola, S., 2013, American Astronomical Society, DPS meeting \#46, id.304.09 
%%
\bibitem[Denk \& Mottola(2015)]{Denk2015}	
Denk, T. \& Mottola, S., 2013, American Astronomical Society, DPS meeting \#47, id.412.02 
%%
\bibitem[Duffard et al.(2009)]{Duffard2009}	
Duffard, R., Ortiz, J.-L., Thirouin, A., Santos-Sanz, P., Morales, N., 2009, A\&A, 505, 1283
%%
\bibitem[Emelyanov et al.(2005)]{Emelyanov}
Emelyanov, N.V., Archinal, B. A., a'Hearn, M. F., et al.,  2005, A\&A, 438, L33 
%%
\bibitem[Engelbracht et al.(2007)]{Engelbracht2007}
Engelbracht, C.W., Blaylock, M., Su, K.Y.L., et al., 2007, PASP, 119, 994
%%
\bibitem[Fruchter \& Hook(2002)]{Fruchter}
Fruchter, A.S., Hook, R.N., 2002, PASP, 114, 144
%%
\bibitem[Gladman et al.(1998)]{Gladman98}
Gladman, B. J., Nicholson, P. D., Burns, J. A., et al., 1998, Nature, 392, 897 
%Discovery of two distant irregular moons of Uranus
%%
\bibitem[Gladman et al.(2000)]{Gladman02}
Gladman, B., Kavelaars, JJ, Holman, M., et al., 2000, Icarus, 147, 320 
%Note: The Discovery of Uranus XIX, XX, and XXI
%%
\bibitem[Gladman et al.(2001)]{Gladman01}
Gladman, B., Kavelaars, J. J., Holman, M., Nicholson, P. D., et al., 2001, Nature, 412, 163
%%
\bibitem[Gomes et al.(2015)]{Gomes15}
Gomes-Júnior, A. R., Assafin, M., Vieira-Martins, R., et al., 2015, A\&A 580, A76 
% Astrometric positions for 18 irregular satellites of giant planets from 23 years of observations (Sycorax)
%%
\bibitem[Gordon et al.(2005)]{Gordon2005}
Gordon, K.D., Rieke, G.H., Engelbracht, C.W., et al., 2005, PASP, 117, 503
%%
\bibitem[Gordon et al.(2007)]{Gordon2007}
Gordon, K.D., Engelbracht, C.W., Fadda, D., et al., 2007, PASP, 119, 1019
%%
%%
\bibitem[Grav et al.(2003)]{Grav03}
Grav, T.,Holman, M. J., Gladman, B. J., Aksnes, K., 2003, Icarus, 166, 33
%Photometric Survey of the Irregular Satellites
%%
\bibitem[Grav et al.(2004)]{Grav04}
Grav, T., Holman, M. J., Fraser, W. C., 2004, ApJ, 613, L77
%Photometry of irregular satellites of Uranus and Neptune
%%
\bibitem[Grav et al.(2015)]{Grav2015}
Grav, T., Bauer, J. M., Mainzer, A. K., et al., 2015, ApJ, 809, 3G 
%NEOWISE: Observations of the Irregular Satellites of Jupiter and Saturn
%%
\bibitem[Harris(1998)]{Harris98}
Harris, A. W., 1998, Icarus, 131, 291
%%
\bibitem[Heppenheimer \& Porco(1977)]{Heppe}
Heppenheimer, T. A., \& Porco, C. 1977, Icarus, 30, 385 %New contributions to the problem of capture
%%
\bibitem[Hicks \& Buratti(2004)]{Hick2004}
Hicks, M.D., Buratti, B.J., 2004, Icarus, 171, 210
%The spectral variability of Triton from 1997–2000 
%%
\bibitem[Holman et al.(2000)]{Holman99}
Holman, M., Gladman, B., Kavelaars, JJ, et al., 2000, DPS 32.4201 
%The discovery and recovery of Uranus XVIII-XX
%%
\bibitem[Holman et al.(2003)]{Holman03}
Holman, M., Kavelaars, J., Grav, T., et al., 2003, IAU Circular 8047
%%
\bibitem[Howell et al.(2014)]{Howell14}
Howell, S.B., Sobeck, C., Haas, M., et al., 2014, PASP, 126, 398
% The K2 Mission: Characterization and Early Results
%%
\bibitem[Karkoschka(2001)]{Karkoschka2001}
Karkoschka, E., 2001, Icarus, 151, 51K
%Comprehensive Photometry of the Rings and 16 Satellites of Uranus with the Hubble Space Telescope
%%
\bibitem[Kavelaars et al.(2004)]{Kavelaars04}
Kavelaars, J. J., Holman, M. J., Grav, T., et al., 2004, Icarus, 169, 474
% The discovery of faint irregular satellites of Uranus
%%
\bibitem[Kiss et al.(2014)]{Kiss2014}
Kiss, Cs., M\"uller, T.G., Vilenius, E., et al., 2014, Experimental Astronomy, 37, 161
%%
\bibitem[Kiss et al.(2016)]{Kiss2016}
Kiss, Cs., P\'al, A., Farkas-Tak\'acs, A., et al., 2016, MNRAS, 457, 2908
%%
\bibitem[Lacerda et al.(2014)]{L14}
Lacerda, P., Fornasier, S., Lellouch, E., et al., 2014, ApJL, 793, L2
%%
\bibitem[Lagerros(1996)]{Lagerros1}
Lagerros, J. S. V., 1996, A\&A, 310, 1011
%%
\bibitem[Lagerros(1997)]{Lagerros2}
Lagerros, J. S. V., 1997, A\&A, 325, 1226
%%
\bibitem[Lagerros(1998)]{Lagerros3}
Lagerros, J. S. V., 1998, A\&A, 332, 1123
%%
\bibitem[Lenz \& Breger(2005)]{Lenz}
Lenz, P., \& Breger, M. 2005, Commun. Asteroseismol., 146, 53
%%
\bibitem[Lellouch et al.(2013)]{Lellouch13}
Lellouch, E., Santos-Sanz, P., Lacerda, P., et al., 2013, A\&A, 557, A60
%%
\bibitem[Luu(1991)]{Luu}
Luu, J., 1991, AJ, 102, 1213
%%
\bibitem[Luu \& Jewitt(1990)]{LJ}
Luu, J. X., Jewitt, D. C., 1990, AJ, 99, 1985
%%
\bibitem[Maris et al.(2001)]{Maris01}
Maris, M., Carraro, G., Cremonese, G., Fulle, M., 2001, AJ, 121, 2800 
%Multicolor photometry of the Uranus irregular satellites Sycorax and Caliban
%%
\bibitem[Maris et al.(2007)]{Maris07}
Maris, M., Carraro, G., Parisi, M. G., 2007, A\&A, 472, 311 
%Light curves and colours of the faint Uranian irregular satellites Sycorax, Prospero, Stephano, Setebos, and Trinculo
%%%%
%%%%%
\bibitem[Millis \& Thompson(1975)]{Millis1975}
Millis, R.L.\& Thompson, D.T., 1975, Icarus,  26, 408
%UBV photometry of the Galilean satellites
%%
\bibitem[Moln\'ar et al.(2017)]{Molnar2017}
Moln\'ar, L., et al., 2017, ApJS, submitted, arXiv:1706.06056
% Main-belt Asteroids in the K2 Uranus Field
%%
\bibitem[Monet et al.(2003)]{USNO}
Monet, D.G., Levine, S.E., Canzian, B., et al., 2003, AJ, 125, 984
%%
\bibitem[Morrison et al.(2009)]{Morrison2009}
Morrison, S.J., Thomas, P.C., Tiscareno, M.S., Burns, J.A., Veverka, J., 2009, Icarus, 204, 262
%Grooves on small saturnian satellites and other objects: Characteristics and significance 
%%
\bibitem[M\"uller \& Lagerros(1998)]{ML98}
M\"uller, T.G., Lagerros, J.S.V., 1998,  A\&A,  338, 340
%%
\bibitem[Muinonen et al.(2010)]{Muinonen2010}
Muinonen, K., Belskaya, I.N., Cellino, A., et al., 2010, Icarus, 209, 542
%%
\bibitem[M\"uller et al.(2009)]{Muller2009} 
M\"uller, Th.G., Lellouch, E. B\"ohnhardt, H, et al.,  Earth, Moon, and Planets, 105, 209-219
%%
\bibitem[\protect\citeauthoryear{Mueller et al.}{2012}]{Migo}
Mueller, M., Stansberry, J., Mommert, M. \& Grundy, W., 2012, "TNO Diameters And Albedos: The Final MIPS Dataset", AAS DPS meeting, \#44, \#310.13
%%
\bibitem[\protect\citeauthoryear{M\"uller \& Lagerros}{2002}]{ML2002}
M\"uller, T. G. \& Lagerros, J. S. V., 2002, A\&A, 381, 324
%%
\bibitem[\protect\citeauthoryear{M\"uller et al.}{2011}]{Muller2011}
M\"uller, T., Okumura, K., Klaas, U., 2011, "PACS Photometer Passbands and Colour Correction Factors for Various Source SEDs", PICC-ME-TN-038 (Herschel/PACS calibration report)
%%
\bibitem[Nicholson et al.(2008)]{Nicholson}
Nicholson, P.D., Cuk, M., Sheppard, S.S., et al., 2008, Irregular Satellites of the Giant Planets, in: The Solar System Beyond Neptune,  p.411 (Tuscon, AZ: Univ. Arizona Press)
%%
\bibitem[P\'al(2012)]{Pal2012}
P\'al, A., 2012, MNRAS, 421, 1825
%% FITSH- a software package for image processing
%
\bibitem[P\'al et al.(2015)]{Pal15}
P\'al, A., Szab\'o, R., Szab\'o, Gy. M., 2015, ApJ, 804, L45
% Pushing the limits: K2 observations of the trans-Neptunian objects 2002 GV31 and (278361) 2007 JJ43
%%
\bibitem[P\'al et al.(2016)]{Pal16}
P\'al, A., Kiss, Cs., Thomas M.G.et al., 2016, AJ, 151, 117
% Large size and slow rotation of the trans-Neptunian object (225088) 2007 OR10 discovered from Herschel and K2 observations
%%
\bibitem[Parisi et al.(2008)]{Parisi08}
Parisi, M. G., Carraro, G., Maris, M., Brunini, A., 2008, A\&A 482, 657
% Constraints to Uranus’ great collision IV The origin of Prospero
%%
\bibitem[Pilcher et al.(2012)]{Pilcher}
Pilcher, F., Mottola, S., Denk, T., 2012, Icarus, 219, 741
%%
\bibitem[Poglitsch et al.(2010)]{PACS}
Poglitsch A. et al., 2010, A\&A, 518, L2
%%
\bibitem[Pollack et al.(1879)]{Pollack}
Pollack, J. B., Burns, J. A., Tauber, M. E., 1979, Icarus, 37, 587 
%Gas drag in primordial circumplanetary envelopes - A mechanism for satellite capture
%%
\bibitem[Porco et al.(2003)]{Porco03}
Porco, C. C., West, R. A., McEwen, A., et al., 2003, Science, 299,  1541
%%
\bibitem[Porco et al.(2005)]{Porco05}
Porco, C. C., et al., 2005, Science, 307, 1237
%%%
\bibitem[Pravec et al.(2002)]{Pravec2002}
Pravec, P., Harris, A. W., Michalowski, T., 2002, Asteroid Rotations, in: Asteroids III, Univ. of Arizona Press. 
%%%
\bibitem[Pravec \& Harris(2000)]{Pravec+Harris}
Pravec, P., \& Harris, A. W., 2000, Icarus, 148, 12
%%%%
\bibitem[Rettig et al.(2001)]{Rettig2001}
Rettig, T. W., Walsh K. \& Consolmagno, G., 2001, Icarus, 154, 313
%Implied Evolutionary Differences of the Jovian Irregular Satellites from a BVR Color Survey 
%%
\bibitem[\protect\citeauthoryear{Rieke et al.}{2004}]{MIPS}
Rieke, G. H., Young, E. T., Engelbracht, C. W., et al., 2004, ApJS, 154, 25
%%
\bibitem[Romon et al.(2001)]{Romon01}
Romon, J., de Bergh, C., Barucci, M. A. et al., 2001, A\&A, 376, 310 
%Photometric and spectroscopic observations of Sycorax, satellite of Uranus
%%
\bibitem[\protect\citeauthoryear{Schaefer et al.}{2008}]{Schaefer2008}
Schaefer, B. E., Tourtellotte, S. W., Rabinowitz, D. L., Schaefer, M. W., 2008, Icarus, 196, 225
%Photometric and spectroscopic observations of Sycorax, satellite of Uranus
%%
\bibitem[Sheppard et al.(2003a)]{Sheppard03}
Sheppard, S. S., Jewitt, D., 2003, Science, 423, 261
%%
\bibitem[Sheppard et al.(2003b)]{SheppardIAUC}
Sheppard, S. S., Gladman, B., Marsden, B. G., 2003, IAU Circular 8116
%%
\bibitem[Sheppard et al.(2005)]{Sheppard05}
Sheppard, S. S., Jewitt, D., Kleyna, J., 2005, AJ, 129, 518 
% An ultradeep survey for irregular satellites of Uranus: limits to completeness 
%%
%%
\bibitem[Sheppard et al.(2006)]{Sheppard06}
Sheppard, S. S., Jewitt, D., Kleyna, J., 2006, AJ, 132, 171 
% 
%%
\bibitem[Showalter(2006)]{Showalter2006}
Showalter M. R., Lissauer, J. J., 2006, Science, 311, 973
%The Second Ring-Moon System of Uranus: Discovery and Dynamics
%%%
%%
\bibitem[Simonelli \& Veverka(1984)]{Simonelli1984}
Simonelli D. P. \& Veverka J., 1984, Icarus, 59, 406
%Voyager disk-integrated photometry of Io
%%%
\bibitem[Smith(1989)]{Smith1989}
Smith, B. A., Soderblom, L. A., Banfield, D., et al., 1989, Science, 246, 1422
%Voyager 2 at Neptune: Imaging Science Results
%%
\bibitem[Stansberry et al.(2007)]{Stansberry2007}
Stansberry, J.A., Gordon, K.D., Bhattacharya, B., et al., 2007, PASP, 119, 1038
%%
\bibitem[\protect\citeauthoryear{Stansberry et al.}{2008}]{stansberry2008} 
Stansberry, J., Grundy, W.M., Brown, M.E., et al., 2008, Physical Properties of Kuiper Belt and Centaur Objects: Constraints from the Spitzer Space Telescope, in: The Solar System Beyond Neptune, p.161 (Tuscon, AZ: Univ. Arizona Press)
%%%%
\bibitem[\protect\citeauthoryear{Stansberry et al.}{2012}]{stansberry2012}
Stansberry, J.A., Grundy, W.M., Mueller, M., et al., 2012, Icarus, 219, 676
%%%%
\bibitem[Szab\'o et al.(2013)]{Szabo2013}
Szab\'o, R., Szab\'o, Gy.M., D\'alya, G., et al., 2013, A\&A, 553, A17 
%%%%%
\bibitem[Szab\'o et al.(2016)]{Szabo2016}
Szab\'o, R., P\'al, A., S\'arneczky, K., et al., 2016, A\&A, 596, A40
%% Uninterrupted optical light curves of main-belt asteroids from the K2 Mission
%
\bibitem[Szab\'o et al.(2017)]{Szabo2017}
Szab\'o, M. Gy., P\'al., A., Kiss, Cs., et al., 2017, A\&A, 599, 44
%% The heart of the swarm: K2 photometry and rotational characteristics of 56 Jovian Trojan asteroids
%%
\bibitem[Thirouin et al.(2014)]{Thirouin2014}
Thirouin, A., Noll, K.S., Ortiz, J.-L., Morales, N., 2014, A\&A, 569, A3
%%
\bibitem[Thomas et al.(1991)]{Thomas1991}
Thomas P., Veverka J., Helfenstein P., 1991, J. Geophys. Res. Suppl., 96, 19253
%%
\bibitem[Vilenius et al.(2014)]{Vilenius2014}
Vilenius, E., Kiss, Cs., M\"uller, Th. G., et al. 2014, A\&A, 564, A35
%%
\end{thebibliography}
\end{document}